**The optimization of crop response to climatic stress through modulation of plant stress response mechanisms. Opportunities for biostimulants and plant hormones to meet climate challenges.**

Jing Li[1]*, Giulia Forghieri[2]*, Danny Geelen[1], Patrick du Jardin[3], Patrick H. Brown[4]#

1 HortiCell, Department Plants and Crops, Faculty of Bioscience Engineering, Ghent University, Coupure Links 653, 9000 Ghent, Belgium

2 CatMat lab, Department of Molecular Sciences and Nanosystems, Ca' Foscari University of Venice and Consortium INSTM UdR VE, via Torino 155, Venice 30172, Italy

3 Plant Sciences, Gembloux Agro-Bio Tech, University of Liège, Gembloux, 4000, Belgium

4 Department of Plant Science, University of California Davis, One Shields Ave, 95616, Davis, CA, USA

*These authors contributed equally.

#Corresponce: phbrown@ucdavis.edu

**Abstract**

This review contrasts the molecular and agronomic effects of the use of breeding techniques and agronomic interventions, including biostimulants, plant hormones and plant nutrients to enhance crop stress resilience to climate stress through the modulation of natural physiological pathways. While the use of plant breeding, plant nutrients or other cropping strategies to improve crop resilience is well accepted and not subject to explicit regulatory oversight, the use of biostimulants or plant hormones to achieve the same goal, even when it occurs through the same physiological processes and involves similar changes in plant growth regulatory networks, is subject to strict regulatory oversight if changes in the regulation of plant growth and development are claimed or implied. Evidence is presented to demonstrate that this inconsistent application of regulatory frameworks is not founded in science and that the use of exogenous products such as biostimulants, plant hormones, or microbial products to positively modulate natural plant growth and development pathways to optimize resilience in the face of stress, is safe and not fundamentally different than achieving the same goal through plant breeding or accepted agronomic practices.

Climate change is a major threat to crop potential and is characterized by both long-term shifts in temperature and precipitation patterns as well as increased occurrence of extreme weather events. These extreme weather events are the most immediate and intractable threat to agriculture. Crop resilience in the face of stress depends upon the speed and effectiveness with which plants and cropping systems sense and respond to that stress. A variety of agronomic practices including breeding, exogenous inputs (nutrients, water, biostimulants and others) and shifts in cultivation practice have been used to influence plant stress response to achieve the goal of increased plant and cropping system resilience. Traditional breeding is a powerful tool that has resulted in stable and long-term cultivar improvements but is often too slow and complex to meet the diverse, complex and unpredictable challenges of climate induced stresses. Increased inputs (water, nutrients, pesticides etc.) and management strategies (cropping system choice, soil management etc.) can alleviate stress but are often constrained by cost and availability of inputs. Exogenous biostimulants, microbials and plant hormones have shown great promise as mechanisms to optimize natural plant resilience resulting in immediate but non-permanent improvements in plant responses to climate induced stresses.

Modern molecular tools have provided a much better understanding of the mechanisms by which plants respond to stress, allowing the development of strategies to enhance systems resilience. When exposed to environmental change, a cascade of signaling processes are orchestrated that ultimately determines plant response. Plant hormones are key regulators of this process, enabling rapid and adaptive stress responses through their biosynthesis, metabolism, signaling, and transport. Plant stress resilience inevitably involves changes in plant hormone pathways as a natural response of plants to their environment. Modern cultivars and cropping systems, however, exhibit diminished climate resilience as a consequence of the loss of these natural response pathways through breeding and selection. Targeted modification of plant stress response mechanisms and plant regulatory networks and hormone mediated pathways, represents a powerful but underexplored approach to enhance crop productivity under climate stress. The use of biostimulants, plant hormones and microbes to improve copping system resilience is however, constrained by outdated and scientifically unsound regulations.

The failure to modernize regulatory frameworks for the use of biostimulants in agriculture will constrain the development of safe effective tools and deprive growers of means to respond to the vagaries of climate change. Here we discuss the scientific rationale for eliminating the regulatory barriers that constrain the potential for biostimulants or products that modulate plant regulatory networks to address climate change challenges and propose a framework for enabling legislation to strengthen cropping system resilience.

# 1 Introduction

The United Nations issued a red alert in 2023 after new records were set for every major climate indicator (WMO, 2024). Recent projections indicate that the effects of climate change will emerge earlier than expected, with several major crop-producing regions likely to experience significant impacts before 2040 (Jägermeyr *et al.*, 2021). By 2050, it is projected that an additional 20% of the global population could face hunger due to the impact of a once-in-100-year extreme climate event (Hasegawa *et al.*, 2021). As erratic and extreme weather patterns intensify, traditional farming systems are becoming increasingly vulnerable.

Changes in average temperature and precipitation affect crop adaptability by influencing photosynthesis, respiration, and water use, while the increased frequency and intensity of climate extremes pose more complex physiological and agronomic challenges. In Brazil, for example, weather variability over the past two decades has caused a 50% increase in yield fluctuations for major crops (Burney *et al.*, 2024). Extreme short-term events—such as frost, heat, drought, and flooding—pose a particular disruption to cropping systems. Examples include wet or cold springs, which delay cereal crop planting, reduce germination and emergence, and shorten the growing season, ultimately lowering yields. Insufficient winter chill disrupts flowering in temperate trees and fruit species, reducing fruit set. Heat spells during flowering impair seed set in many crops. Erratic rainfall and unusual heat patterns extend and weaken flowering in tropical species like coffee, resulting in uneven ripening, increased harvesting costs, and reduced yield and quality. These extreme events not only affect crop physiology but also disrupt routine farming practices, adding costs, risks, and reduced profitability to growers.

To cope with climate stress, native species have evolved highly sophisticated adaptive plasticity that enables them to respond effectively to environmental changes (Brooker *et al.*, 2022). Plant adaptive plasticity refers to the physiological mechanisms that allow plants to adjust to growth-limiting resource shortages in variable environments. Many modern crop species, however, are significantly less tolerant to climate variability compared to their wild relatives (Landis *et al.*, 2024). The focus on selecting a limited number of high-yielding, commercially valuable cultivars has led to crop genetic erosion, diminishing both the adaptive plasticity and genetic diversity necessary to cope with climate change (Khoury *et al.*, 2022). This, coupled with decreased plant species diversity in modern cropping systems, excessive use of some agricultural chemicals and degradation of soil health has resulted in loss of plant and soil microbiome diversity and further undermined cropping system resilience (He *et al.*, 2020).

Adaptive plasticity is in part mediated through plant hormones. Plant hormones are signaling molecules that regulate physiological processes and developmental programs in response to both endogenous signals and environmental cues (Lichtfouse, 2021). The biosynthesis of all plant hormones originates from central metabolite precursors, creating intricate interactions among sugars, intermediates of the tricarboxylic acid cycle, ascorbate, inositol phosphate, and hormones (Fàbregas and Fernie, 2021). Extensive crosstalk among plant hormones establishes a complex regulatory network that fine-tunes the balance between growth and stress responses.

To compensate for the reduced adaptive plasticity of modern cropping systems, technologies are needed that that enhance crop resilience. The centrality of plant hormone pathways in crop stress resilience suggests that the targeted modification of plant hormone response networks will be a critical strategy in the development of crops and cropping systems with greater resilience to the threat of climate induced crop stress. Modification of plant hormone response networks to improve crop stress response has historically been achieved through breeding and through increased agronomic inputs (irrigation, fertilization, crop protection etc.) and more recently through the application of plant biostimulants, plant hormones and microbial products. Each approach has distinct benefits and constraints.

Crop breeding is a strategy that has long been employed to improve abiotic stress tolerance, though with mixed success. Breeding for stress tolerance is complex since the impacts of climate change on crop productivity are variable, multifactor, often localized and cropping system-specific and as a result are slow and challenging. Extreme weather events induced by climate change are unpredictable and often highly localized, and as such extremely difficult to select or breed for. Breeding for stress tolerance is further hindered by our limited understanding of the genetic mechanisms underlying these traits, while the uncertainty of climate change-induced disruptions makes trait selection difficult.

Climate stress tolerance can also be managed through increased crop inputs (water, nutrients, soil amendments), and through improved management technologies, such as conservation tillage, shading, and frost prevention, each of which can help mitigate environmental stress. While these approaches can be effective, rapid and flexible, they are time intensive and depend upon the availability and cost effectiveness of the needed inputs.

The application of exogenous chemicals including microbial products (Zhang *et al.*, 2021), biofertilizers and biostimulants for the management of plant stress responses is an area of tremendous interest and unmet potential that has the added benefit of being rapid and targeted with generally low cost of implementation. The use of biostimulants and plant hormones to achieve climate stress tolerance is however, strongly constrained by lack of understanding of the mechanisms involved (Walia, 2023) and by regulatory restrictions that constrain the use of any product that explicitly targets plant growth and development processes. This constraint applies even if the changes in plant growth and development or plant hormone levels that result from product application do not differ from those that occur naturally in well adapted species. Significantly, crop breeding, plant nutrient additions or the use of certain microbial inoculants that also directly alter plant growth and development, and cause changes in plant hormone response networks, do not face the same regulatory constraints as biostimulants or plant growth regulators even when identical outcomes are achieved. These practices are widely considered safe.

This review examines the mechanism of plant response to the environment and tolerance to climate stress, contrasts three strategies to address climate stress and highlights both the similarities, opportunities and constraints of these approaches. The implication of these approaches for the development of sound regulatory frameworks governing the use of plant hormones and biostimulants in sustainable agriculture, is discussed.

## 2 The Mechanisms of Plant Stress Response

Plants respond to abiotic stress through a complex cascade of signaling events, starting with perception and signal transduction, followed by the induction of stress-related genes and downstream processes **(Figure 1A)**. Early environmental signals are sensed by plants or their microbial partners, which then are converted into chemical messages that transduce from the cellular level to the organ and ultimately to the whole plant (Zhang *et al.*, 2021). Upon stress perception, rapid changes occur in plant response pathways and their regulatory networks, including second messengers, transcriptional reprogramming, transcript processing, and post-translational protein modifications. This regulatory signaling network also governs primary and secondary metabolism, such as plant hormones (Sulpice and McKeown, 2015). Our understanding of plant stress response pathways is complicated by the nonlinear nature of responses across environmental gradients and phenological stages (Arnold *et al.*, 2019). Moreover, epigenetic mechanisms that enable plants to adapt through priming and stress-dependent memory add complexity, creating a temporal disconnect between the observed stress response and the current stress condition (Gallusci *et al.*, 2023).

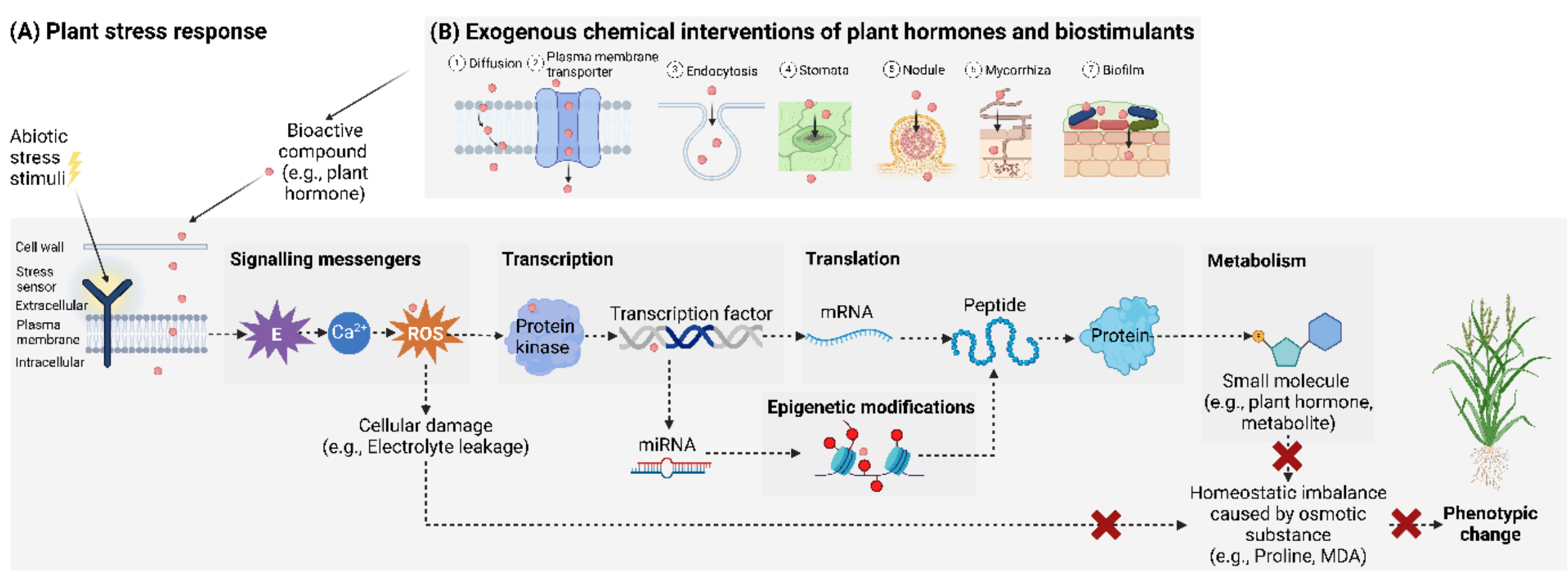

**Figure 1.** The mechanisms underlying plant stress responses (A) and the roles of exogenous bioactive compounds in modulating cascade pathways (B). Plant hormones and biostimulants are examples of bioactive compounds that can be either derived from natural substances or chemically synthesized. (E: electrical signal; $Ca^{2+}$: calcium ions; ROS: reactive oxygen species; mRNA: messenger RNA; MDA: malondialdehyde. )

Plant cells and their subcellular compartments have specialized sensors or sensory systems that detect early deviations from homeostasis in response to stress (Dietz and Vogelsang, 2024). During signal transduction, spikes and waves of electrical signals and secondary messengers, such as $Ca^{2+}$ and reactive oxygen species (ROS, such as $H_2O_2$), function as signaling molecules under normal conditions. However, their overproduction leads to oxidative stress, causing low sodium potassium ratio ($Na^+/Ka^+$), ultimately electrolyte leakage (EL) and cellular damage (Turkan, 2018). Proline accumulates as an osmotic substance during drought stress, helping sustain photosynthetic electron transport (Alvarez *et al.*, 2022). Similarly, free malondialdehyde (MDA) levels rise under stress conditions, induced by ROS or lipoxygenase activity (Alvarez *et al.*, 2022). Non-enzymatic antioxidants scavenge free radicals or indirectly regulate housekeeping enzymes to mitigate oxidative damage (Mittler *et al.*, 2022). Downstream protein kinases, regulated via central metabolism, coordinate the growth-defense tradeoff by managing resource allocation (He *et al.*, 2022). These include well-characterized pathways such as the Sucrose non-fermentable 1-related protein kinase 1 (SnRK1) and the target of rapamycin (TOR) pathways (Baena-González and Hanson, 2017). Stress-related gene expression can also be regulated through transcription factors that further modulate translation processes, resulting in altered production of peptides, proteins, and small molecules (Chen and Koehler, 2020). Additionally, epigenetic modifications, including DNA methylation, chromatin remodeling, synthesis of regulatory small RNA molecules, and histone dynamics, may also occur (Chang *et al.*, 2020). Central metabolism mediates endogenous hormone biosynthesis and regulates other functions—such as metabolism, perception, signaling, and transport—at the intersection of plant stress responses (Fàbregas and Fernie, 2021).

## 3 Plant Hormones in Plant Stress Response

The central role of plant hormones in stress perception, signaling and responses suggest that there is potential to target the modification of hormone signaling pathways for improved resilience. In plants, cellular activity is fine-tuned through local adjustments in hormone levels, which are dynamically regulated by biosynthesis, catabolism, transport, and signal perception **(Figure 2)**. The active forms of plant hormones are synthesized from precursor molecules derived from primary metabolites, such as amino acids and nucleotides, or converted from more complex secondary metabolites through specific biosynthetic and catabolic enzymes.

Hormone homeostasis is further regulated by conjugation with sugars or amino acids, often, but not always, resulting in inactive hormone derivatives.

Plant cells express various hormone receptors that trigger physiological responses, such as proton pump activation, or initiate downstream signaling to regulate gene transcription. Natural plant hormones are highly mobile, moving between cells through the apoplast via efflux and influx carrier proteins or through plasmodesmata connecting adjacent cells. This mobility allows hormones to be transported between tissues and organs, such as from roots to shoots. Additionally, hormones can travel long distances through vascular tissues, influencing their spatial and temporal activity. Each level of hormone transport and signaling presents opportunities for targeted modification using genetic tools or exogenous bioactive compounds to fine-tune plant responses.

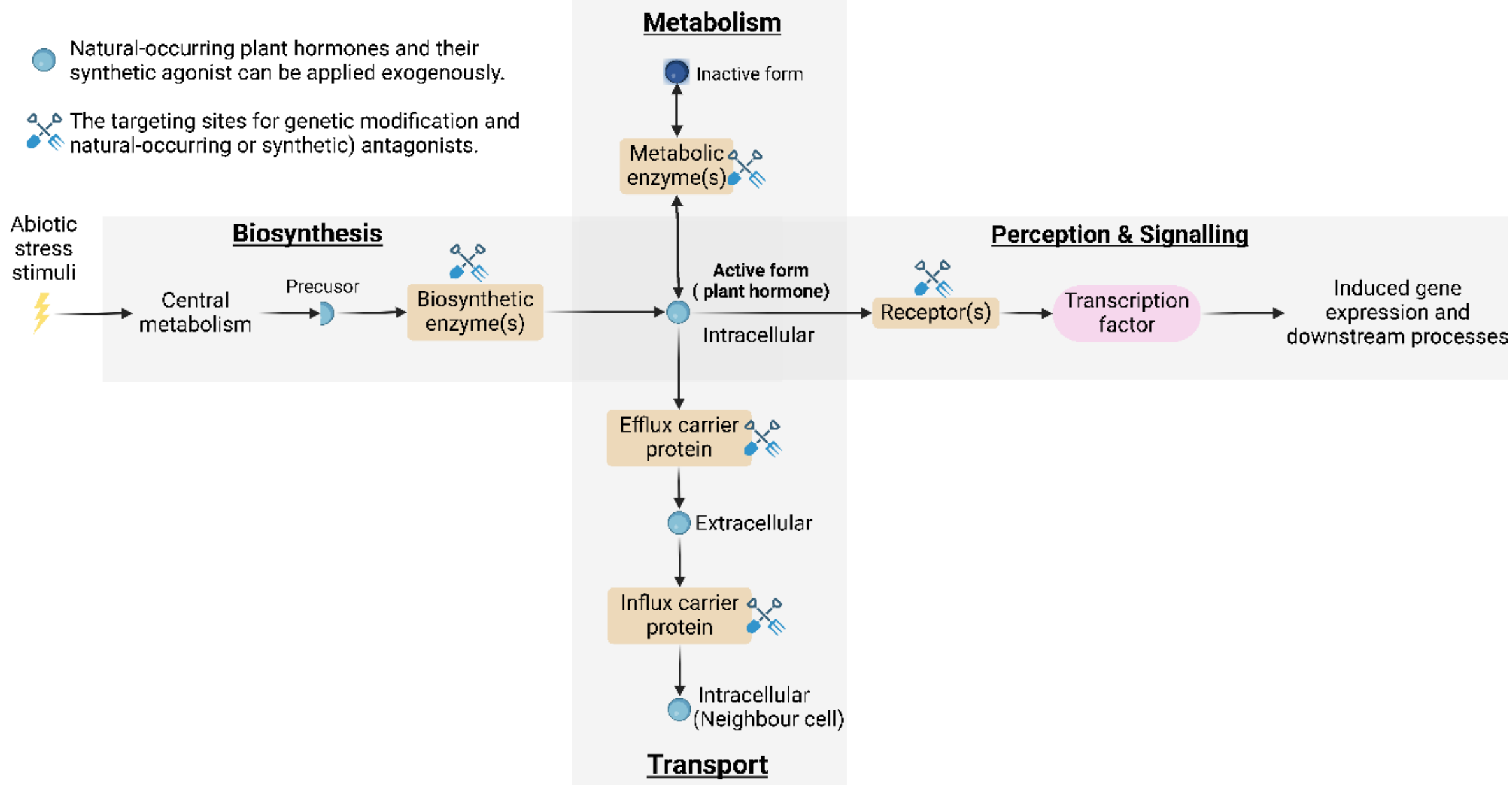

**Figure 2.** Hormone homeostasis is mediated through biosynthesis, metabolism, transport, signaling and perception.

The study of plant hormone functions in growth and development is a major topic in plant biological research. Auxin supports meristem maintenance, stomatal dynamics, and tropic growth responses (Zhao, 2018). Abscisic acid is synthesized under stress, inducing stomatal closure to enhance water-use efficiency (Kuromori *et al.*, 2018). Brassinosteroids regulate cell elongation and stem cell differentiation in the root apex (Oklestkova *et al.*, 2015). Cytokinins control cell fate decisions, improving plant adaptability (Zwack and Rashotte, 2015). Ethylene mediates stress responses and influences cell division and expansion contextually (Dubois *et al.*, 2018). Gibberellins coordinate nitrogen-responsive growth regulation (Wu *et al.*, 2021). Jasmonic acid modulates seedling growth, leaf senescence, and reproductive processes (Huang *et al.*, 2017). Salicylic acid accumulates under stress, inducing growth suppression mechanisms (Pokotylo *et al.*, 2022). Strigolactones regulate plant architecture and impact rhizosphere interactions (Chesterfield *et al.*, 2020).

### 3.1 Genetic Approaches To enhance Plant Stress Tolerance Through Hormone Pathway Modifications

Over the past decades, extensive efforts have been made to engineer hormone signaling pathways through breeding approaches, yet examples of commercial implementation are limited. Mutation breeding based on selective advantage has proven highly effective, leading to the official release of over 3,220 mutant cultivars across more than 210 plant species (Bado *et al.*, 2015). Breeding strategies integrate forward and reverse genetics to identify and characterize genes responsible for mutant phenotypes while analyzing the effects of gene disruptions on plant traits (Jankowicz-Cieslak and Till, 2015). Single-nucleotide polymorphisms (SNPs) linked to desirable crop traits are identified through quantitative trait locus (QTL) mapping, genome-wide association studies (GWAS), whole-genome resequencing (Meyer and Purugganan, 2013), and more recently, pan-genome analyses spanning landraces, cultivars, and wild crop species (Jayakodi *et al.*, 2020).

A diverse array of mutants modulating plant hormone pathways have been identified, targeting biosynthetic and metabolic enzymes, receptors, and transporters **(Figure 2)**. **Table 1** summarizes single-allele loss-of-function (reduced or abolished protein activity) and gain-of-function mutations (enhanced protein activity) that alter hormone pathways and influence crop phenotypic changes under stress conditions. The effects of these mutations on stress tolerance vary on a case-by-case basis. For example, OsGH3-2 overexpression in rice, which encodes an enzyme catalyzing IAA (a natural auxin) conjugation, enhances cold tolerance but increases drought sensitivity due to its modulation of both IAA and abscisic acid homeostasis (Du *et al.*, 2012). Similarly, cytokinin-deficient potato lines overexpressing AtCKX2 (Arabidopsis thaliana CYTOKININ OXIDASE/DEHYDROGENASE2) exhibit improved tolerance to mild salinity stress but reduced tolerance to severe salinity stress compared to the wild type (Raspor *et al.*, 2024).

Despite growing evidence supporting the role of hormone signaling mutants in stress tolerance, their practical application remains limited (Li and Chen, 2023). Several challenges hinder implementation. Most mutations in hormone pathways arise from random variations induced by physical or chemical mutagenesis rather than natural biological processes such as domestication or diversification (Bado *et al.*, 2015). Hormone signaling and stress response networks involve numerous genes with complex interactions, making it difficult to modify a single gene without unintended negative consequences. Even with a well-characterized target protein, mutational scanning is necessary to evaluate effects on activity, stability, dosage, and potential pleiotropic outcomes (Soskine and Tawfik, 2010). While CRISPR/Cas has improved precision genome editing, concerns remain regarding off-target effects (Manghwar *et al.*, 2019). In summary, genetic modifications of hormone pathways hold promise for breeding but remain costly, technically challenging, and fraught with risks.

**Table 1.** Molecular manipulation of plant hormone pathways and their influence on crop stress responses.

| Plant hormone | Mutant variation | Treated crop | Stress condition | Mechanism involved | Genetic technique and target gene | Phenotype changes |
|---|---|---|---|---|---|---|
| Auxin | Loss-of-function | Tomato | Drought | Signaling | Deficient *AFB4* gene encoding Auxin-signaling F-Box protein | Vegetative growth↑(Yu *et al.*, 2015) |
| | | | Salt | | | Vegetative growth↑(Bouzroud *et al.*, 2020) |
| | | Tomato | Cadmium | | Deficient auxin-insensitive DIAGEOTROPICA (DGT) gene | Vegetative growth↑(Alves *et al.*, 2017) |
| | | Rice | Heat | | Deficient *IAA29* gene encoding AUX/IAA protein | Grain yield ↓(Chen *et al.*, 2024) |
| | | Rice | Aluminum | Transport | Deficient *AUX3* gene encoding AUXIN1/LIKE AUX1 protein | Seedling growth ↑(Wang *et al.*, 2019*a*) |
| | | Rice | Cadmium | | Deficient *AUX1* gene encoding AUXIN1/LIKE AUX1 protein | Seedling growth ↓(Yu *et al.*, 2015) |
| | | Rice | Cold and drought | | Deficient *PIN1b* gene encoding PIN protein | Vegetative growth ↓(Yang *et al.*, 2023) |
| | Gain-of-function | Rice | Drought | Biosynthesis | Overexpressing *GH3* gene encoding IAA-amido synthetase | Seedling growth ↓(Du *et al.*, 2012) |
| | | | Cold | | | Seedling growth ↑(Du *et al.*, 2012) |
| Abscisic acid | Loss-of-function | Tomato | Drought | Biosynthesis | Deficient *notabilis/flacca (not/flc)* double mutant that frameshift mutation in the NCED1 gene | Fruit size↓(Nitsch *et al.*, 2012) |
| | | Barley | Drought | | Deficient Az34 mutant (*nar2a* gene) encoding molybdopterin | Seedling growth↓(Walker-Simmons *et al.*, 1989) |
| | | Tomato | Drought | | Deficient *Sitiens* gene encoding dehydrogenase | Vegetative growth↓(Nitsch *et al.*, 2012) |
| | | Tomato | Drought or salt | | Deficient *flacca* gene encoding dehydrogenase | Vegetative growth↓(Grillo *et al.*, 1995) |
| | | Potato | Drought | | Sensitive mutant blocked at ABA-aldehyde | Vegetative growth↓(Etehadnia *et al.*, 2008) |

| | | Tobacco | Drought | | Deficient *aba1* gene encoding zeaxanthin epoxidase | Vegetative growth↑(Mizokami *et al.*, 2015) |
|---|---|---|---|---|---|---|
| | Gain-of-function | Tomato | Salt | Biosynthesis | Overexpressing the NCED1 gene encoding enzyme 9-cis-epoxycarotenoid dioxygenase | Vegetative growth↑, fruit yield↑(Martínez-Andújar *et al.*, 2021) |
| | | Tobacco | Drought | | | Vegetative growth↑(Qin and Zeevaart, 2002) |
| Brassinosteroid | Loss-of-function | Tomato | Heat | Signaling | Insensitive curl3$^{-abs}$ gene | Seedling growth ↓(Mazorra *et al.*, 2011) |
| | Gain-of-function | Tomato | Heat | Biosynthesis | Overexpressing *Dwarf* gene | Seedling growth ↑(Mazorra *et al.*, 2011) |
| | | Oilseed | Drought or heat | | Overexpressing *DWF4* gene encoding C-22 hydroxylase | Vegetative growth↑, grain yield↑(Sahni *et al.*, 2016) |
| | | Tomato | Cold | Signaling | Overexpressing BRI1 gene encoding BR receptor kinase | Vegetative growth↑(Wang *et al.*, 2022*a*) |
| | | Maize | Salt | | Overexpressing *BSK1* gene encoding BR-signaling kinase | Vegetative growth↑(Liu *et al.*, 2022) |
| | | Creeping bentgrass | Drought | Metabolism | Overexpressing *BAT1* gene encoding BR-related acyltransferase | Vegetative growth↑(Han *et al.*, 2017) |
| Cytokine | Loss-of-function | Tomato | Drought or heat | Signaling | Downregulating *HK2* gene encoding histidine kinase | Vegetative growth↑(Mushtaq *et al.*, 2022) |
| | | Tomato | Salt | | Downregulating *AHP* gene encoding histidine phosphotransfer protein | Seedling growth↓(Sun *et al.*, 2014) |
| | Gain-of-function | Creeping bentgrass | Heat | Biosynthesis | Overexpressing *ipt* gene encoding isopentenyltransferase | Vegetative growth↑(Xu *et al.*, 2009) |
| | | | Drought | | | Vegetative growth↑(Xu *et al.*, 2016) |
| | | Barley | Drought | Metabolism | Overexpressing *CKX1* gene encoding cytokinin dehydrogenase | Vegetative growth↑(Pospíšilová *et al.*, 2016) |
| | | Tobacco | Drought | | | Vegetative growth↑(Lubovská *et al.*, 2014) |
| | | Potato | Salt (mild) | | Overexpressing *CKX2* gene encoding cytokinin dehydrogenase 1 | Seedling growth↑(Raspor *et al.*, 2024) |
| | | | Salt (severe) | | | Seedling growth↓(Raspor *et al.*, 2024) |
| | | Rice | Drought or salt | | Overexpressing *LOG* gene encoding phosphoribohydrolase | Vegetative growth↑, grain yield ↑(Rathore *et al.*, 2024) |
| Ethylene | Loss-of-function | Tobacco | Salt | Biosynthesis | Deficient ACS gene encoding 1-aminocyclopropane-1-carboxylate synthase | Vegetative growth↑(Wi *et al.*, 2010) |
| | Gain-of-function | Rice | Drought and | Biosynthesis | Overexpressing ET overproducer gene ETOL1 | Vegetative growth↑(Du *et al.*, 2014) |
| | | | flooding | | | Vegetative growth↓(Du *et al.*, 2014) |
| | | Tobacco | Salt | Signaling | Overexpressing *TERF1* gene encoding ethylene response factor protein | Vegetative growth↑(Tian *et al.*, 2011) |
| | | Tomato | Salt | | | Vegetative growth↑(Huang *et al.*, 2004) |
| | | Rice | Drought | | | Vegetative growth↑(Zhang *et al.*, 2010) |
| | | Rice | Cold | | Overexpressing *TERF2* gene encoding ethylene response factor protein | Vegetative growth↑(Du *et al.*, 2014) |
| Gibberellin | Loss-of-function | Rice | Cold | Biosynthesis | Deficient GA-insensitive dwarf1 gene | Vegetative growth ↑ (Tanaka *et al.*, 2006) |
| | | Sunflower | Drought | | Deficient GA-insensitive *dwarf2* gene encoding ent-kaurenoic acid oxidase | Vegetative growth↑(Mariotti *et al.*, 2022) |
| | | Tomato | Drought | | Deficient *gid1* or *gid2* gene | Vegetative growth↑(Omena-Garcia *et al.*, 2019) |
| | | Maize | Drought | | Deficient *ks3-1* gene encoding kaurene synthase | Vegetative growth↑(Wu *et al.*, 2023) |
| | Gain-of-function | Rice | Salt | Biosynthesis | Overexpressing *GA2ox5* gene encoding GA 2-oxidase | Vegetative growth↑(Wu *et al.*, 2023) |
| Jasmonic acid | Loss-of-function | Tomato | Salt | Biosynthesis | Deficient *def-1* gene encoding defenseless-1 | Vegetative growth↑(Abouelsaad and Renault, 2018) |
| | | Tomato | Cadmium | | Deficient spr2 gene encoding chloroplast fatty acid desaturase | Vegetative growth↓(Zhao *et al.*, 2016) |
| | | Maize | Salt | | Deficient *opr7opr8* gene encoding oxo-phytodienoic acid reductases | Seedling growth↓(Ahmad *et al.*, 2019) |
| | Gain-of-function | Rice | Drought | Metabolism | Overexpressing *JMT* gene encoding jasmonic acid carboxyl methyltransferase | Reproductive growth↓, grain yield ↓(Kim *et al.*, 2009) |
| Salicylic acid | Loss-of-function | Tobacco | Drought | Signaling | Overexpressing SABP gene encoding SA binding protein 2 | Vegetative growth↑(Li *et al.*, 2019) |
| Strigolactone | Loss-of-function | Barley | Drought | Signaling | Deficient *hvd14.d* gene encoding α/β hydrolase | Vegetative growth↓(Marzec *et al.*, 2020) |
| | | Tomato | Drought | | Downregulating *CCD7* gene encoding carotenoid-cleavage dioxygenases | Vegetative growth↓(Visentin *et al.*, 2016) |

Gene name in *italic*.

### 3.2 The Roles of Exogenous Hormones in Crop Stress Regulation

Exogenous bioactive compounds, whether naturally occurring or chemically synthesized, may help alleviate plant stress by modulating various pathways, minimizing cellular damage, restoring homeostasis and supporting growth maintenance **(Figure 1B)**. These compounds can be applied directly or induced through microbial inoculation. Plant uptake can occur via multiple pathways, including passive diffusion, active transport through plasma membrane transporters, endocytosis, stomata, nodulation, mycorrhiza, and biofilms. The following section presents examples of how various hormones, hormone analogs, and inhibitors influence plant stress tolerance. **Box 1** summarizes classical hormones and analogs involved in plant growth and development, while **Table 2** illustrates how the exogenous application of these compounds, including both agonists and antagonists, enhances stress tolerance across different crop species.

**Box 1. Glossary of terms (in the context of this review)**

Active form: A hormone in an active form that is able to cause a specific effect.

Agonist: A hormone analog that mimics the mode of action of a naturally occurring hormone.

Analog: A small molecule whose physical structure is similar to a naturally occurring hormone.

Antagonist: A hormone analog that blocks the mode of action of a naturally occurring hormone.

Inactive form: A hormone in an inactive form that fails to cause a specific effect.

Precursor: The starting compound that is used for hormone biosynthesis.

Naturally occurring auxins, including indole-3-acetic acid (IAA), 4-chloroindole-3-acetic acid, indole-3-butyric acid (IBA), and phenylacetic acid (PAA), have been utilized to manipulate plant stress responses. For instance, the exogenous application of IAA improved grain yield in rice under drought and heat stress by enhancing pollen viability and spikelet fertility (Sharma *et al.*, 2018). Similarly, rapeseed seeds soaked in IBA exhibited significantly enhanced protective mechanisms and resilience in vegetative growth under salinity stress (Li *et al.*, 2024). Additionally, a wide range of synthetic auxins has been developed to over-activate or inhibit auxin signaling, leading to applications in weed control and defoliation (Busi *et al.*, 2018). While synthetic and natural auxins are primarily used for growth promotion or inhibition, auxin also plays a crucial role in regulating crop stress responses (Weijers *et al.*, 2018). For example, the application of naphthaleneacetic acid (NAA) induces early IAA-dependent accumulation of $H_2O_2$, enhancing antioxidant capacity and improving drought stress tolerance in soybean seedlings (Xing *et al.*, 2016*a*).

Abscisic acid occurs as the (+)-cis,trans-isomer (S-ABA), a chiral structure that is challenging to synthesize and is commercially produced through the fermentation of phytopathogenic fungi (Rademacher, 2015). The application of S-ABA has been shown to enhance stress tolerance in various crops. For example, in salt-treated rice, S-ABA improved photosynthesis by reducing $Na^+$ accumulation in leaves and increasing the activity of antioxidant enzymes (Jiang *et al.*, 2024). Similar stress-alleviating effects have been observed in maize and apple under drought conditions (Tworkoski *et al.*, 2011; Yao *et al.*, 2019; Qiao *et al.*, 2023). Additionally, synthetic receptor agonists of abscisic acid, such as pyrabactin analogs (e.g., Quinabactin and Opabactin), have been developed to selectively control stomatal conductance, offering potential applications for reducing water loss during drought stress (Liu *et al.*, 2023). For instance, Quinabactin treatment reduced water loss from detached leaves in soybean, barley, and maize, ultimately increasing final biomass (Okamoto *et al.*, 2013).

Brassinosteroids include approximately 70 polyhydroxylated sterol derivatives, such as 24-epibrassinolide (EBL), which is found in 64 different plant species (Oklestkova *et al.*, 2015). Despite significant efforts to chemically synthesize brassinosteroids, only two brassinosteroid analogs, EBL and BB16, are widely used in agriculture (Oklestkova *et al.*, 2015). Foliar applications of both EBL and BB16 have been shown to enhance fruit production and improve fruit quality in strawberry plants subjected to salt and water stress (Furio *et al.*, 2022). Additionally, these two brassinosteroids have been validated in independent studies for their protective effects on other crops such as rice and lettuce (Núñez Vázquez *et al.*, 2013; Serna *et al.*, 2015; Reyes Guerrero *et al.*, 2017).

Natural cytokines are $N^6$-substituted adenine derivatives, including compounds such as N,N'-diphenylurea, $N^6$-[(3-methylbut-2-en-1-yl)amino]purine (iP), zeatin, Z-9-riboside, and meta-Topolin, all of which isolated from plant materials (Kamínek, 2015). While many synthetic 6-benzyladenine (6-BA)-containing compounds with CK-like bioactivities have been reported, only a few are commercially available due to challenges in achieving efficient synthesis (Rademacher, 2015). Exogenous application of 6-BA has been shown to alleviate waterlogging-induced damage in maize by activating the ROS scavenging system and promoting plant growth (Wang *et al.*, 2021). Additionally, two novel cytokine derivatives, PI-55 and INCYDE, have been developed to block cytokine signaling and inhibit cytokines degradation by targeting cytokine oxidase/dehydrogenase. These treatments prolong cytokine activity in plants and have been shown to positively affect shoot and root growth, as well as fresh weight, in medicinal plant seedlings grown in the presence of cadmium (Gemrotová *et al.*, 2013).

Ethylene applications are commonly used to promote rapid and uniform ripening in fruits (Brecht, 2019). Soluble solid forms of ethylene analogs, which either activate ethylene signaling or inhibit ethylene biosynthetic enzymes, are more convenient for plant use and transportation (Depaepe *et al.*, 2020). For example, mustard plants treated with ethephon, an ethylene agonist, under salt stress showed a significant reduction in oxidative damage and an enhancement in photosynthetic nitrogen use efficiency (Iqbal *et al.*, 2017). On the other hand, aminoethoxyvinylglycine (AVG) and 1-methylcyclopropene (1-MCP) are two effective ethylene antagonists that inhibit both ethylene biosynthesis and signaling pathways. Treatments with AVG and 1-MCP prevented weight loss and reduced the decline in soluble solids content during cold storage of persimmon fruit by minimizing the incidence and severity of peel blackening and softening (Win *et al.*, 2021). Moreover, AVG and 1-MCP also function as stress protectors during the growth period of crops such as rice, cotton, and apple (Djanaguiraman *et al.*, 2011; Chen *et al.*, 2014; Mohammed *et al.*, 2015).

Nearly 140 different gibberellins have been identified in higher plants and fungi. Commercially, gibberellins are typically obtained through the fermentation of phytopathogenic fungi, with $GA_3$ being the most widely produced form, followed by $GA_4$ and $GA_7$ (Rademacher, 2015). The application of $GA_3$ has been shown to improve grain yield in salt-stressed wheat, with the effect being particularly pronounced in salt-sensitive cultivars (Iqbal and Ashraf, 2013). Several gibberellin biosynthesis inhibitors act as growth retardants by reducing longitudinal shoot growth, resulting in a more compact plant architecture that lowers the risk of lodging and improves tolerance to various abiotic stresses. For example, strawberries treated with Paclobutrazol (PBZ) under drought conditions exhibited enhanced enzymatic and non-enzymatic antioxidant activities, increased relative water content, and improved photosynthetic rate, resulting in higher fruit yield (Saleem *et al.*, 2024).

Among jasmonates, only (+)-7-iso-JA-L-isoleucine (JA-Ile), a stereoisomer conjugated with the amino acid isoleucine, has been recognized for its bioactivity in plants upon exogenous application (Yan *et al.*, 2016). The oxidative damage caused by salt stress in black locust was alleviated by the application of MeJA, which significantly enhanced the activities of antioxidant enzymes (Jiang *et al.*, 2016). Crops such as beans, cauliflower and okra have improved resistance to various stresses through the application of MeJA (Wu *et al.*, 2012; Mohi-Ud-Din *et al.*, 2021; Wang *et al.*, 2023*b*,*a*).

Natural salicylic acid is primarily produced by plants, and its analogs are widely used as pharmaceuticals due to their anti-inflammatory properties (Rosheen *et al.*, 2023). Salicylic acid or acetylsalicylic acid (ASA) treatments, applied through seed soaking or root treatment, significantly enhanced the tolerance of maize seedlings and young plants to chilling stress (Wang *et al.*, 2012). Other stress-tolerant properties of salicylic acid have also been observed in crops like rice, tomato, and sunflower (Jini and Joseph, 2017; Noreen *et al.*, 2017; Khan *et al.*, 2019; Naeem *et al.*, 2020). However, most synthetic salicylic acid analogs have primarily shown to improve plant immunity against biotic stress, with limited evidence supporting their role in abiotic stress tolerance (Tripathi *et al.*, 2019).

Two stereochemical families of strigolactones, namely (+)-strigol and (−)-orobanchol, are the most abundant in nature. Their complex structures make large-scale production challenging. Many studies on synthetic strigolactone analogs easier to synthesize structures have primarily focused on their herbicidal activity, while research on their role in abiotic stress response has been comparatively limited (Zwanenburg and Blanco-Ania, 2018). GR24, the most widely used synthetic strigolactone, has protective effects against abiotic stresses in barley, orange, cucumber, maize, and cotton (Qiu *et al.*, 2021; Ma *et al.*, 2022*a*,*b*; Zhang *et al.*, 2022; Luqman *et al.*, 2023; Song *et al.*, 2023).

Although the commercialization outlook for hormone-like molecules is promising, with an increasing number of newly developed analogs demonstrating greater potency, stability, and efficacy in agricultural applications, their use in agriculture remains limited. Several factors contribute to this: insufficient studies on a variety of crops, a lack of long-term and detailed dose-response trials, and a generally insufficient understanding of the metabolic processes involved. These limitations have hindered the commercial adoption of these products to date. Furthermore, the relatively limited commercial use of natural and synthetic plant hormones to manage plant stress responses is also due to regulatory constraints, as products that alter plant growth regulatory processes are often classified as pesticides, which severely limits their commercial development (EPA, 2024).

**Table 2.** The application of natural-occurring plant hormones and their synthetic analogs and their influence on crop stress responses.

| Plant hormone | Chemical nature | Compound name | Function | Application method | Treated crop | Stress condition | ROS | Osmotic substance | Mitigated phenotype |
|---|---|---|---|---|---|---|---|---|---|
| Abscisic acid | Natural occurring | (+)-cis,trans-ABA (S-ABA) | Signaling | Foliar spray pretreatment | Maize | Drought | ↑ | Proline↑, MDA↓ | Germination↑, seedling growth ↑ (Yan *et al.*, 2016) |
| | | | | Foliar spray | Rice | Salt | ↓ | $Na^+/K^+$↓ | Vegetative growth↑ (Jiang *et al.*, 2024) |
| | Synthetic agonist | Quinabactin | Signaling | Foliar spray pretreatment | Soybean, barley, maize | Drought | - | - | Vegetative growth↑ (Okamoto *et al.*, 2013) |
| | | Opabactin | | Foliar spray pretreatment | Wheat | Drought | - | - | Vegetative growth↑ (Vaidya *et al.*, 2019) |
| Auxin | Natural occurring | Indole-3-acetic acid (IAA) | Signaling | Foliar spray pretreatment | Rice | Drought and heat | ↓ | - | Grain yield↑ (Sharma *et al.*, 2018) |
| | Synthetic antagonist | Naphthaleneacetic acid (NAA) | Signaling | Foliar spray | Chufa | Alkaline | ↓ | $Na^+/K^+$↓, MDA↓, | Seedling growth ↑ (Ullah *et al.*, 2022) |
| | | | | Soaking pretreatment | Pea | Drought | ↓ | MDA↓, EL↓ | Seedling growth ↑ (Xing *et al.*, 2016*b*) |
| | | 2,4-dichlorophenoxyacetic acid (2,4-D) | | Soaking pretreatment | Wheat | Salt | ↓ | $Na^+/K^+$↓, MDA↓, EL↓ | Seedling growth ↑ (Mohsin *et al.*, 2020) |
| Brassinosteroid | Natural occurring | 24-epibrassinolide (EBL) | Signaling | Foliar spray | Wheat | Drought | ↓ | Proline↓, MDA↓,EL↓ | Vegetative growth↑, grain yield↑ (Khan *et al.*, 2021*a*) |
| | | | | Foliar spray pretreatment | Perennial Ryegrass | Salt | - | $Na^+/K^+$↓, Proline ↑ | Vegetative growth↑ (Wu *et al.*, 2017) |
| | Synthetic agonist | BB16 | | Foliar spray pretreatment | Strawberry | Salt, drought | - | - | Vegetative growth↑, fruit yield↑, fruit quality↑ (Furio *et al.*, 2022) |
| Cytokinin | Synthetic agonist | 6-Benzyladenine (6-BA) | Signaling | Foliar spray | Maize | Flooding | ↓ | MDA↓, EL↓ | Seedling growth↑ (Wang *et al.*, 2021) |
| | | | | Foliar spray | Winter wheat | Heat | - | - | Reproductive growth↑ (Yang *et al.*, 2016) |
| | | Kinetin | | Foliar spray | Common sage | Salt | - | $Na^+/K^+$↓ | Vegetative growth↑(Tounekti *et al.*, 2011) |
| | | N-(2-chloro-4-pyridyl)-N′-phenylurea (CPPU) | | Foliar spray | Rice | Drought | - | - | Vegetative growth↑ (Gujjar *et al.*, 2020) |
| | | | | Foliar spray pretreatment | Rice | Salt | - | Proline↑ | Vegetative growth↑, grain yield↑ (Gashaw *et al.*, 2014) |
| | Synthetic antagonist | PI-55 | Signaling | Seed treatment | *Bulbine natalensis* Baker *and Rumex crispus* L. | Cadmium | - | - | Seedling growth↑ (Gemrotová *et al.*, 2013) |
| | | 2-chloro-6-(3-methoxyphenyl)aminopurine (INCYDE) | Metabolism | | | | | | |
| Ethylene | Synthetic agonist | Ethephon | Signaling | Foliar spray | Mustard | Salt | ↓ | Proline↑ | Vegetative growth↑ (Iqbal *et al.*, 2017) |

| | Synthetic antagonist | Aminoethoxyvinylglycine (AVG) | Biosynthesis | Foliar spray pretreatment | Cotton | Flooding | - | - | Vegetative growth↑, fruit yield↑ (Najeeb *et al.*, 2015) |
|---|---|---|---|---|---|---|---|---|---|
| | | | | Dipping pretreatment | Persimmon | Cold storage | | | ruit quality↑ (Win *et al.*, 2021) |
| | | 1-Methylcyclopropene (1-MCP) | Signaling | Irrigation | Rice | Salt | ↓ | Proline↓, MDA↓ | Vegetative growth↑, grain yield↑ (Hussain *et al.*, 2019) |
| Gibberellin | Natural occurring | $GA_3$ | Signaling | Foliar spray | Mustard | Salt | - | $Na^+/K^+$↓, MDA↓, EL↓ | Vegetative growth↑ (Siddiqui *et al.*, 2008) |
| | | | | Seed pre-soaking | Wheat | Salt | - | $Na^+/K^+$↓ | Vegetative growth↑, grain yield↑ (Iqbal and Ashraf, 2013) |
| | Synthetic antagonist | Chlormequat chloride (CCC) | Biosynthesis | Seed treatment | Rice | Salt | - | - | Grain yield↑ (Gurmani *et al.*, 2011) |
| | | Mepiquat chloride (MC) | | Seed priming | Cotton | Salt | - | $Na^+/K^+$↓ | Seedling growth↑(Wang *et al.*, 2019*b*) |
| | | Flurprimidol | | Soil drench pretreatment | Red Firespike | Drought | - | - | Vegetative growth↑ (Rezazadeh *et al.*, 2016) |
| | | Trinexapac-ethyl | | Foliar spray pretreatment | Perennial ryegrass | Drought | - | Proline↑, MDA↓, EL↓ | Vegetative growth↑ (Sheikh Mohammadi *et al.*, 2017) |
| | | | | Foliar spray pretreatment | Kentucky bluegrass | Drought | - | EL↓ | Vegetative growth↑ (Krishnan and Merewitz, 2015) |
| | | Paclobutrazol (PBZ) | | Foliar spray or soil drench pretreatment | Pomegranate | Cold | - | Proline↓, EL↓ | Vegetative growth↑ (Moradi *et al.*, 2017) |
| | | | | Foliar spray | Wheat | Heat stress due to late Sow | ↓ | MDA↓,EL↓ | Vegetative growth↑, grain yield↑ (Nagar *et al.*, 2021) |
| | | | | Foliar spray | Strawberry | Drought | ↓ | Proline↓ | Vegetative growth↑, fruit yield↑ (Saleem *et al.*, 2024) |
| Jasmonic acid | Agonist | Methyl jasmonate (MeJA) | Signaling | Soaking | Black locust | Salt | ↓ | $Na^+$↓, MDA↓ | Vegetative growth↑ (Jiang *et al.*, 2016) |
| | | | | Vapor incubation pretreatment | Peach | Cold storage | ↓ | $Na^+$↓, MDA↓ | Postharvest fruit quality↑(Jin *et al.*, 2013) |
| Salicylic acid | Natural occurring | 2-Hydroxybenzoic acid | Signaling | Foliar spray | Maize | Drought | ↓ | MDA↓, | Vegetative growth ↑ (Saruhan *et al.*, 2012) |
| | | | | Foliar spray | Mustard | Salt | ↓ | $Na^+$↓ | Vegetative growth ↑ (Nazara *et al.*, 2015) |
| | | Acetylsalicylic acid (ASA) | | Seed soaking pretreatment and root treatment | Maize | Cold | ↓ | MDA↓ | Seedling growth↑ (Wang *et al.*, 2012) |
| Strigolactone | Agonist | GR24 | Signaling | Foliar spray pretreatment | Cucumber | Salt | ↓ | Proline↑, EL↓, $Na^+/K^+$↓ | Seedling growth↑ (Zhang *et al.*, 2022) |
| | | | | Root treatment | Barley | Cadmium | ↓ | MDA↓ | Vegetative growth↑ (Qiu *et al.*, 2021) |
| | | | | Foliar spray | Maize | Drought | | MDA↓ | Vegetative growth↑, grain yield↑ (Luqman *et al.*, 2023) |
| | | | | Foliar spray | Apple | Saline-alkali | ↓ | $Na^+/K^+$↓ | Seedling growth↑ (Ma *et al.*, 2022*a*) |
| | | | | Dipping pretreatment | Sweet orange | Cold storage | ↓ | MDA↓ | Fruit quality↑ (Ma *et al.*, 2022*b*) |
| | | | | Foliar spray pretreatment | Cotton | Salt | ↓ | Proline↓, MDA↓ | Seedling growth↑ (Song *et al.*, 2023) |

-: not reported; ROS: reactive oxygen species, including hydrogen peroxide ($H_2O_2$), superoxide radicals ($O^{2-}$), and hydroxyl free radical ($OH^-$); $Na^+$: sodium ions; EL: Electrolyte leakage; MDA: Malondialdehyde.

### 3.3 Biostimulants and their Role in Abiotic Stress Tolerance

Biostimulants have emerged as promising products to enhance crop resilience against abiotic stresses (Franzoni *et al.*, 2022, Preprint). According to the International Standards Organization definition, biostimulants are "product(s) that contain substance(s), microorganism(s), or mixtures thereof, that, when applied to seeds, plants, the rhizosphere, soil, or other growth media, act to support a plant's natural nutrition processes independently of the biostimulant's nutrient content" (ISO). Biostimulants can contain various natural compounds of microbial or non-microbial origin, as well as beneficial microorganisms, including bacteria, fungi and yeasts. The most widely studied biostimulants include those derived from seaweed, humic substances, protein hydrolysates, and living microbes. A recent meta-analysis assessed the effectiveness of various biostimulants on crop yield improvement and estimated an increase of about 18% (Li *et al.*, 2022, Preprint). Biostimulants also offer the potential for rapid and broad-spectrum solutions designed to alleviate imminent, ongoing, or recently experienced stress events to crops. **Table 3** summarizes some examples of different biostimulants and their role in plant stress mitigation. Here we highlight some of the studies that provide experimental evidence for a direct mechanistic effect and discuss the possible mode of actions of the different compounds.

#### 3.3.1 Non-microbial Biostimulants

Seaweed-derived biostimulants, like seaweed extracts (SE), contain a rich array of bioactive compounds, including phytohormones like auxins, CKs, and GAs, vitamins, amino acids, and polysaccharides, which contribute to their beneficial effects (Nanda *et al.*, 2021). Tomato treated with SE improved fruit yield under salinity conditions (Hernández-Herrera *et al.*, 2022). The effect was partially attributed to the presence of phytohormones in the extract. The plant hormones auxin, GAs and CKs have been detected in SE and can persist in the final applied extract although in variable concentrations (Sangha *et al.*, 2014). For example, GAs and CKs have been observed to range from 0.3 to4.7 µg $g^{-1}$ and from 0.06 to 4.6 µg $g^{-1}$ of seaweed dry weight, respectively, while auxin have been identified with concentrations between 0.01 and 12 µg $g^{-1}$. SE have shown the capability to alter the endogenous levels of plant hormones (Deolu-Ajayi *et al.*, 2022, Preprint). SE from *Kappaphycus alvarezii* significantly increased endogenous ABA and CKs concentrations in durum wheat in both non-stressed and drought conditions (Patel *et al.*, 2018). The observed increase of maize tolerance to drought stress upon SE application was attributed to both the increase of polyamines endogenous levels and to the stimulation of IAA and GA endogenous production, contributing to the alleviation of stress-induced growth inhibition (Li *et al.*, 2018).

Many SE biostimulants enhance stress resistance even though they do not contain physiologically relevant concentrations of hormones or hormone agonists by boosting the plant's antioxidant defenses, by triggering various physiological and biochemical pathways, such as the synthesis of osmoprotectants and thus reducing oxidative stress caused by abiotic factors (Ali *et al.*, 2022*b*). The stress-protective effect of SE against salt stress was associated with the presence of phenolic compounds that can both act as ROS scavengers and chelate toxic ions (Carillo et al., 2020). Antioxidant and stress-protective effects have also been attributed to the presence of non-structural carbohydrates and biopolymers (Elansary *et al.*, 2016). Algal polysaccharides such as ulvans, alginates and fucans can act as elicitors by activating stress-related pathways in plants, thus resulting in hormone-like effects (Chanda *et al.*, 2019, Preprint). In-soil application of algal polysaccharides improved salt stress tolerance of wheat, as function of the molecular weight and of the sulfate content of the polymers (Bouteraa *et al.*, 2022), and induced an antioxidant response by modulating $Na^+$ uptake and mobilization within the crop and by regulating the expression of $Na^+$ transporters (Zou *et al.*, 2019). Finally, the presence of polyamines in SE can have osmoregulatory effects and protect crops in water deficiency conditions.

Plant extracts (PE) are usually concentrated liquids or powders extracted from various plant species. Some PE may contain natural plant growth regulators, such as auxins, CKs, and Gas, which is an inevitable consequence of their plant origin. PE’s can also induce hormone production. The application of PE on common flax increased the endogenous levels of $GA_3$ and IAA both at the shooting and rooting stages (Oguz, 2024). Other bioactive compounds present in PE, such as sugars, can up-regulate growth-related genes meanwhile downregulating stress-related ones, and thus alleviate salt stress in rice (Ho *et al.*, 2001). Extracts obtained from *Moringa oleifera* protected common bean from salinity and heat stress (Latif and Mohamed, 2016*a*). PE also contain a variety of other bioactive compounds, including alkaloids, flavonoids, phenols, terpenoids, and essential oils, which contribute to their biological properties.

Fulvic acids (FA), humic acids (HA), and humates are highly recalcitrant organic substances derived from the decomposition of plant and animal matter. They are increasingly recognized for their growth-promoting properties, particularly in enhancing crop resilience to abiotic stress. Foliar application of HA improved drought resistance of wheat by increasing its antioxidant response (Arslan *et al.*). The stress-protective effects of HA and humates were attributed to a hormone-like activity leading to the activation of stress-dependent pathways (Olaetxea *et al.*, 2018; Canellas *et al.*, 2024, Preprint). HA can help regulate osmotic pressure within plant cells, aiding in the maintenance of turgor pressure and overall cellular function during water stress, and improving grain yield by 16% in two different maize genotypes exposed to drought stress (Chen *et al.*, 2022*b*). These HA-induced stress-protective effects were observed along with shifts in endogenous hormonal levels, including an increase of IAA and a decrease of ABA concentrations. The growth-stimulating effect of HA on wheat was also related to the modulation of endogenous hormonal pathways, as gene analysis revealed the HA-induced up-regulation of genes involved in biosynthesis of auxin and CKs (Rathor *et al.*, 2024).

Soil treatment with HA and FA can improve soil structure by enhancing aeration and water retention. This can lead to better root development and increased access to water and nutrients, and was shown to be particularly beneficial during drought in maize under semi-arid conditions (Zhou *et al.*, 2019). HA and FA were also observed to affect nutrient assimilation by plants and to modulate the expression of antioxidant compounds during abiotic stress periods (Canellas *et al.*, 2015, Preprint), thereby boosting the plant's antioxidant systems and reducing oxidative stress caused by environmental factors. Additionally, humic substances applied to soil can chelate ions and thus limit plants' exposure to toxic cations (Canellas *et al.*, 2015, Preprint). These effects, together with decreased membrane leakage, were associated with HA-mediated alleviation of salt stress on bean (Adil Aydin, 2012).

Protein hydrolysates (PH) are derived from the enzymatic or chemical hydrolysis of proteins from various sources, including animal by-products, plant biomass and microbial fermentations of various carbon compounds (Colla *et al.*, 2014; Gao *et al.*, 2021, Preprint). PH are usually rich in amino acids and peptides, which act as biostimulants through both soil and foliar application (Colla *et al.*, 2015, Preprint). PH stimulate root and shoot development and promote overall plant vigor, which is crucial for coping with abiotic stressors like drought and salinity (Casadesús *et al.*, 2020). In some cases, the presence of tryptophan – an auxin precursor – in PH has been attributed to the stimulation of auxin-like responses resulting in the promotion of seed germination and plant growth of pea (*Pisum sativum* L.) (Colla *et al.*, 2017, Preprint). Root growth promotion by PH can enhance the plant's ability to access water and nutrients, improving resilience against drought and nutrient deficiencies (Casadesús *et al.*, 2020; Ceccarelli *et al.*, 2021). The foliar application of vegetal PH promoted root development by increasing the root number and length of tomato. The PH treatment of water-stressed tomato plants was associated with increased $GA_1$, $GA_3$ and IAA endogenous levels and decreased ABA concentrations (Casadesús *et al.*, 2019). Metabolic analyses suggested that PH of various origin alter both phytohormone profile and fatty acid metabolism of the treated plants (Ceccarelli *et al.*, 2021). PH can improve nutrient solubility and availability, facilitating the uptake of essential nutrients by plants during stress conditions (Rouphael *et al.*, 2020). Seed treatment with PH protected tomato plants from heat and drought-induced damage by preserving yield and quality traits (Francesca *et al.*, 2022). These protective effects were observed along with an increase in antioxidant content within the plant (Wang *et al.*, 2022*b*). The application of PH to maize seedlings in a hydroponic system improved the plant's tolerance to salt, nutrient deficiency and hypoxia stress conditions (Trevisan *et al.*, 2019). The effects were traced back to a PH-related modulation of nitrate transporters and ROS gene expression. PH function can also vary depending on the source of hydrolysates. For example, 2 out of 11 PH tested as seed primers resulted in salt stress alleviating properties (Sorrentino *et al.*, 2021). In contrast, different types of PH had similar but crop-dependent salt stress alleviating effects (Zuluaga *et al.*, 2023).

### 3.3.2 Microbial Biostimulants

Interest in the use of plant microbial inoculation to enhance crop stress tolerance is founded on the observation that the imposition of abiotic stress often results in functional changes to the plant microbiome that enhance plant stress tolerance, and from the observation that stress tolerant species are often associated with specific microbial partners critical for the tolerance of those species to the stress (Timm et al., 2018). Abiotic stress has been shown to modify the plant microbiome by changing the composition of microbial communities in the rhizosphere, and by generally enriching species with higher resistance and ability to adapt to the stress conditions (Fitzpatrick *et al.*). The inoculation of stress-tolerant microbes to the plant has also been shown to increase abiotic stress tolerance (Enebe and Babalola, 2018, Preprint). The microbiome shifts under plant stress vary with the specific abiotic stress conditions, with some "core species" remaining preserved, indicating a strict relationship between the plant and its associated bacteria (Timm *et al.*, 2018). For example, drought conditions enriched Actinobacteria over other Bacteroidetes and Proteobacteria in bulk soil, due to the higher resistance of Actinobacteria to drought conditions. Plant exudates were observed to increase during drought, suggesting that plant-beneficial microbes can be selectively enriched to promote plant tolerance to stress (Omae and Tsuda, 2022). Microbial biostimulants, including both bacteria and fungi, have the potential to synthesize IAA, as predicted from the analysis of 7,282 prokaryotic genomes and empirical evidence (Keswani *et al.*, 2020). The beneficial effect of microbial biostimulants has been often attributed to the production of phytohormones (Ali et al., 2022). It has therefore been widely hypothesized that treating plants with stress-tolerant-plant-beneficial microbes can help restore stress-induced microbiome imbalances and contribute to stress alleviation. Some examples of microbial biostimulants, including bacterial and fungal formulations, are reported in **Table 3**.

The *in-vitro* inoculation of rice seedlings with the rhizobacterium *Bacillus altitudinis* resulted in a phenotypic modification of root architecture that was attributed to a change in IAA endogenous levels within the root and to the genetic modulation of auxin-responsive genes involved in root formation (Ambreetha *et al.*, 2018). A similar increase of endogenous IAA was observed upon the inoculation of wheat with rhizobacteria, including *Dietzia natronolimnaea*, *Arthrobacter protophormiae* and *Bacillus subtilis*, resulting in an increased tolerance against both drought and salinity (Barnawal *et al.*, 2017). A bacterial consortium composed of *Staphylococcus epidermidis* CK9 strain and *Bacillus australimaris* CK11 inoculated on Arabian balsam tree (*Commiphora gileadensis*), improved tolerance to both salinity and drought stress and decreased endogenous levels of ABA and JA, while stimulating SA accumulation (Jan *et al.*, 2024). Further investigations suggested that microbial biostimulants can influence in the expression of the *TaCTR1* gene, involved in plant response to various stress types (Bi *et al.*, 2010). The inoculation of *Bacillus casamancensis* MKS-6 and *Bacillus* sp. MRD-17 in mustard counteracted drought stress by influencing both plant endogenous hormonal levels – including ABA and GAs – and ABA-independent signaling, by downregulating *BjDREB1_2* and *BjDREB2* transcription factors (Nivetha *et al.*, 2024). The presence of bacterial phytohormones has been shown to counteract the growth-inhibition effect of abiotic stress.

Fungal biostimulants have also been effective in plant stress mitigation. Seed pre-treatment with the fungal strain *Trichoderma lixii* improved plant and root development and osmolytes accumulation of maize exposed to salt stress (Pehlivan *et al.*, 2017). It was proposed that *Trichoderma* can both trigger various stress defense responses in plants and stimulate plant rooting and growth parameters via the production of IAA, as well as adsorb and chelate toxic ions via the production of siderophores, including excess $Na^+$ (Colla *et al.*, 2015; Yadav *et al.*, 2024). In durum wheat exposed to drought and salinity stress in a greenhouse experiment, plants treated with a fungal rich microbial formulation were more tolerant compared to untreated plants (Yaghoubi Khanghahi *et al.*, 2022). Results suggested that fungal biostimulant inoculation also improved grain quality of wheat, enhancing protein, sugars, and lipid content under stress.

In addition to the direct effects of microbial biostimulants on phytohormone production, a diversity of microbial-dependent stress-protective mechanisms have been proposed (Table 3). Microbial strains producing 1-aminocyclopropane-1-carboxylate (ACC) deaminase can lower ethylene levels and improve plant response to various types of stresses (Glick, 2014; Jha *et al.*, 2021, Preprint). The multiple inoculation of basil (*Ocimum sanctum* L.) with ACC deaminase-producing microbes resulted in lower levels of ACC, increased yield and phenolic content and consequently alleviated cold-dependent growth inhibition (Singh *et al.*, 2020). Similarly, the inoculation of groundnut (*Arachis hypogea*) exposed to saline stress conditions with ACC-producing *Pseudomonas fluorescent* increased yield (Saravanakumar and Samiyappan, 2007).

The application of microbial biostimulants can result in the release of microbial compounds including extracellular polymeric substances, organic acids, siderophores, glycolipids, polysaccharide and lipoproteins, that can act as chelating agents or bind ions, thereby reducing the mobility of toxic metals in soil (Barra Caracciolo and Terenzi, 2021, Preprint; Ali *et al.*, 2022, Preprint). Various microbial biostimulants have been observed to alleviate plant stress induced by heavy metal pollution, and their consequent accumulation in plants' tissues (Etesami, 2018, Preprint). For instance, the amelioration of $Cd^{2+}$ phytotoxicity in mung bean seedlings inoculated with *Enterobacter asburiae* was attributed to the bacterial production of organic acids (Kavita *et al.*, 2008). In addition, the inoculation of *Azotobacter* sp. in Cd and Cr contaminated soils reduced the uptake of heavy metals by wheat (Joshi and Juwarkar, 2009).

The difficulty in obtaining effective root colonization and persistence in soil and plant over time are challenging issues when assessing the effectiveness of MBS and plant-beneficial microbes (Romano *et al.*, 2020). Efficient delivery systems together with improved understanding of plant-microbe interaction can help reduce yield losses due to abiotic stress and improve the efficacy of MBS in the target plants. The integration of MBS in agricultural practices presents a promising strategy for enhancing plant resilience to abiotic stress, offering potential benefits for crop quality and for agricultural sustainability in changing environmental conditions.

## 4 Effects of the Exogenous Application of Plant Hormones, Plant Nutrients and Biostimulants on Endogenous Hormone Levels and Plant Response.

The extent to which in-vivo plant hormone levels can change under stress and non-stress conditions, during plant development, and in response to the application of hormones, nutrients and biostimulants under different agronomic treatments and in different plant species is provided in **Supplementary Table 1**. Changes in *in-vivo* hormone concentrations reflect both changes in synthesis and metabolism and the activity of regulatory feedback mechanisms that control hormone homeostasis. Accurately measuring the effects of various stress mitigation strategies on in-vivo plant hormone activity will be critical to optimizing the use of exogenous hormones and biostimulants as stress mitigation strategies. Measurement of plant hormones in-vivo is complicated by sensor sensitivity and expression levels show strong temporal and spatial variation (Jones, 2016).

A wealth of literature demonstrates that micro- and macro-nutrients can regulate hormone biosynthesis, signaling, and transport (Rubio et al., 2009; Jia et al., 2022), and nutrient deficiencies can alter the plant's hormonal response, triggering hormonal stress-like responses (Wittenmayer and Merbach, 2005). The application of silicon, for example, has been observed to both correct nutrient deficiency and to enhance plant stress tolerance (Hosseini *et al.*, 2017; Réthoré *et al.*, 2020). Similarly, biostimulants such as humates, have been observed to trigger hormone mediated stress-alleviating effects under nutrient deficient conditions (Othibeng *et al.*, 2021), suggesting that the stress-related response mechanisms, plant nutrient status and related metabolic networks impacting hormonal pathways are often intertwined and not easy to distinguish (Supplementary Table 1). Stress events, including nutrient deficiency or excess, also result in direct changes to cellular hormone levels depending on growth conditions and the species analyzed. Moreover, in response to stress, hormone levels are observed to rise or decrease with effects on primary and secondary metabolism and growth, making it difficult to disentangle the impact of plant hormone levels from other metabolic changes (Supplementary Table 1).

Overall, the level of alteration in endogenous hormone concentrations in response to stress or exogenous application of bio-active compounds, such as biostimulants, nutrients, plant hormones or microbes, is usually small. On the other hand, stress imposition often results in a more consistent change of endogenous hormonal levels, reflecting the role of plant hormones in crop response strategies. Across all stress events and all exogenous products or nutrient applications, the measured changes in internal hormone levels did not result in internal hormone concentrations exceeding 1/100$^{th}$ of the EU Codex Maximum Residue Limit (MRL's) demonstrating that the use of these approaches to address crop stress is safe.

**Table 3.** Effect of the application of non-microbial and microbial biostimulants on crop stress responses.

| **Biostimulant category** | **Natural source** | **Active ingredient** | **Application method** | **Treated crop** | **Stress condition** | **ROS** | **Osmotic substance** | **Mitigated phenotype** |
|---|---|---|---|---|---|---|---|---|
| Seaweed extracts | *Ascophyllum nodosum* | IAA, GA, CK, phenols, biopolymers, sugars | Foliar spray | Tomato | Drought | - | Proline↑ | Fruit yield↑, fruit quality↑ (Ahmed *et al.*, 2022) |
| | *Sargassum spp.* | IAA, phenols, biopolymers, sugars | Foliar spray | Tomato | Salt | ↓ | proline↑ | vegetative growth↑ (Sariñana-Aldaco *et al.*, 2022) |
| | *Sargassum wightii* | IAA, GA, CK, phenols, biopolymers, sugars | Foliar spray | Okra | Salt | ↓ | $Na^+$↓ | Vegetative growth↑, fruit yield↑, fruit quality↑ (Khan *et al.*, 2022) |
| Garlic extract | *Allium sativum* | IAA, GA, ABA, Kinetin, ascorbic acid, sugars | Seed pre-treatment | Broad bean | Drought | ↓ | MDA↓ | Vegetative growth↑ (Kasim, 2017) |
| Carrot extract | *Daucus carota* | IAA, GA, ABA, CK, ascorbic acid, sugars | | | | | | |
| Moringa extracts | Moringa oleifera | IAA, GB, CK, ABA, sugars, phenols, ascorbic acid | Foliar spray | Common bean | Salt, heat | ↓ | MDA↓ | Green pod yield↑(Latif and Mohamed, 2016*b*) |
| Protein hydrolysate | Sugar cane molasses and | Glycine betaine, | Fertigation | Tomato | Drought | ↓ | - | Fruit yield↑ (Francesca *et al.*, 2021) |

| | yeast extract (*Saccharomyces cerevisiae*) | peptides, amino acids | | | | | | |
|---|---|---|---|---|---|---|---|---|
| | Pumpkin seeds | Amino acids, peptides | Foliar spray | Common bean | Salt | ↓ | Proline↓, MDA↓ | Green pod yield↑, Vegetative growth↑ (Sitohy *et al.*, 2020) |
| Humic acid | Leonardite | Humic acid | Soil treatment | Finger millet | Salt | ↓ | Proline↓, MDA↓ | Vegetative growth↑ (Rakkammal *et al.*, 2024) |
| | Non-specified | Humic acid | Seed priming | Rice | Salt | ↓ | $Na^+$↓, MDA↓, Proline↓ | Vegetative growth↑ (Shukry *et al.*, 2023) |
| | Organic matter | Humic acid | Soil treatment | Maize | Drought | - | Proline↑ | Grain yield↑ (Chen *et al.*, 2022*b*) |
| PGPB | *Azotobacter vinellandii* SRI Az3 | IAA, GA | Root inoculation pretreatment | Rice | Drought | - | MDA↓, Proline↑ | Vegetative growth↑ (Pradhan *et al.*, 2018) |
| | *Bacillus amyloliquefaciens RWL-1* | ABA | Root inoculation | Rice | Salt | - | Proline↑ | Seedling growth↑(Shahzad *et al.*, 2017) |
| | *Bacillus cereus* SA1 | IAA, GA | Root inoculation pretreatment | Tomato | Drought | - | $K^+$↑ | Vegetative growth↑ (Khan *et al.*, 2020) |
| | *Bacillus* strains | IAA | Root inoculation pretreatment | Rice | Drought | - | $Na^+/K^+$↓ | Seedling growth↑ (Khan *et al.*, 2021*b*) |
| | *Ensifer meliloti* RD64 | IAA | Seed inoculation pretreatment | Alfalfa | Drought | - | Proline↑ | Vegetative growth↑ (Khan *et al.*, 2021*b*) |
| | *Enterobacter cloacae* | ACC deaminase | Seed pre-treatment | Wheat | Salt, heavy metal | ↓ | MDA↓ | Vegetative growth↑ (Singh *et al.*, 2022) |

| | *Leclercia adecarboxylata* MO1 | IAA | Root inoculation | Tomato | Salt | - | Proline↑ | Vegetative growth↑ (Kang *et al.*, 2019) |
|---|---|---|---|---|---|---|---|---|
| | *Pseudomonas azotoformans* | ET | Root inoculation pretreatment | Tomato | Salt | | Proline↓ | Vegetative growth↑ (Liu *et al.*, 2021) |
| | *Pseudomonas fluorescens* G20-18 | CK | Root inoculation pretreatment | Tomato | Drought | - | - | Vegetative growth↑ (Mekureyaw *et al.*, 2022) |
| | *Pseudomonas putida* H-2-3 | GA | Root inoculation pretreatment | Soybean | Drought | ↓ | $Na^+$↓ | Vegetative growth↑ (Kang *et al.*, 2014) |
| | *Pseudomonas sp.* UW4 | ET | Root inoculation pretreatment | Tomato | Salt | - | - | Vegetative growth↑ (Orozco-Mosqueda *et al.*, 2019) |
| | *Streptomyces sp.*, *Pseudomonas sp.* | Polysaccharide | Seed pre-treatment | Wheat | Salt | - | Proline↓ | Vegetative growth↑ (Thakur and Yadav, 2024) |
| AMF | *Rhizophagus irregularis* | IAA, CK, GA, ET | Root inoculation pretreatment | Black locust | Drought | ↓ | MDA↓ | Vegetative growth↑ (He *et al.*, 2017) |
| | *Funneliformis mosseae* | Biopolymers, chelating compounds | Seed treatment | Maize | Heat | ↓ | Proline↑ | Vegetative growth↑ (Ye *et al.*, 2019) |
| Endophytic fungi | *Paecilomyces formosus* LHL10, *Penicillium funiculosum* LHL06 | Chelating compound | Root inoculation pretreatment | Soybean | Drought, heavy metal | ↓ | MDA↓ | Vegetative growth↑ (Bilal *et al.*, 2020) |
| | *Paecilomyces formosus* LHL10 | GA, IAA | Root inoculation pretreatment | Cucumber | Salt | ↓ | MDA↓ | Vegetative growth↑ (Latif Khan *et al.*, 2012) |
| | *Trichoderma spp.* | GA, ABA, SA, IAA, CK | Root inoculation pretreatment | Wheat | Drought | ↓ | - | Vegetative growth↑ (Illescas *et al.*, 2021) |

Species are indicated in *italic.* PGPB: plant growth-promoting bacteria; AMF: Arbuscular mycorrhizal fungi; -: not reported; ROS: reactive oxygen species, including hydrogen peroxide ($H_2O_2$), superoxide radicals ($O^{2-}$), and hydroxyl free radical ($OH^-$); $Na^+$: sodium ions; $Cl^-$: chloride ions; EL: Electrolyte leakage; MDA: Malondialdehyde.

**Supplemental Table 1.** Endogenous levels of auxin (IAA), abscisic acid (ABA), cytokinins (CK), gibberellins (GA), jasmonic acid (JA) and salicylic acid (SA) expressed as percent (%) increase or decrease under stress conditions and in response to application of exogenous compounds and treatments. Data expressed as a percentage change from control.

| **Compound** | **Experimental conditions** | **Stress** | **Crop** | **% Increase (+) decrease (–)** | | | | | | **Ref.** |
|---|---|---|---|---|---|---|---|---|---|---|
| | | | | IAA | ABA | CK | GA | JA | SA | |
| None | Growth chamber | Salinity | Wheat | -13 | +57 | -50 | | | | (Shakirova *et al.*) |
| None | Hydroponic system | Sulfur deficiency | Rice leaves | | +135 | | | +13 | +92 | (Réthoré et al., 2020) |
| None | Growth chamber | Cold | Wheat spikelet | -55 | +27 | | -41 | | | (Zhang *et al.*, 2019) |
| None | Growth chamber | Drought | Maize leaves | -60 | +40 | | -70 | | | (Wang *et al.*, 2008) |
| None | Greenhouse pot | Salinity | Soybean | | +62 | | | +15 | +37 | (Hamayun *et al.*, 2010) |
| None | Growth chamber | Drought | Rice leaves | -30 | | | | +43 | | (Du *et al.*, 2013) |
| | | Cold | | +43 | | | | +48 | | |
| | | Heat | | -13 | | | | -34 | | |
| Manganese | Greenhouse pots | Salt | Rice shoot | +21 | | | | | | (Nadeem et al., 2020) |
| Zinc | Greenhouse pots | Salt | Rice shoot | +34 | -10 | | | | | |
| Copper sulfate | Hydroponic system | None | Maize leaves | +279 | +126 | | +29 | +100 | -17 | (Reckova et al., 2019) |
| Hydrogen sulphide | Growth chamber | Iron deficiency | Soybean leaves | +62 | -34 | | +32 | | +17 | (Chen et al., 2020) |

| Iron | Growth chamber | Iron deficiency | Soybean shoots | +20 | -20 | | +80 | | +20 | |
|---|---|---|---|---|---|---|---|---|---|---|
| Molybdenum | Growth chamber | Cold | Wheat | | –43 | | | | | (Sun *et al.*, 2009) |
| Phosphorous | Hydroponic system | Nutrient deficiency | Barley roots | –50 | +175 | +450 | | +6.7 | –45 | (Nadira *et al.*, 2016) |
| Silicon | Hydroponic system | Salt | Soybean | | –24 | | | | | (Lee *et al.*, 2010) |
| Silicon | Growth chamber | Potassium deficiency and osmotic | Barley leaves | | +100 | +47 | | | | (Hosseini et al., 2017) |
| SA | Growth chamber | Salinity | Wheat | | +35 | | | | | (Shakirova *et al.*) |
| SA | Growth chamber | None | Wheat | +12 | +14 | +8 | | | | (Shakirova *et al.*) |
| $GA_3$ | Greenhouse pot | Salinity | Soybean | | -63 | | +77 | +20 | -17 | (Hamayun *et al.*, 2010) |
| IAA | Field | None | Fragrant Pear calyx tube | +88 | –8.3 | | –10 | | | (Chen *et al.*, 2022*a*) |
| Naphthalene acetic acid | Growth chamber | Drought | Soybean | +20 | +53 | | | | | (Xing *et al.*, 2016*b*) |
| SA | Growth chamber | Salt | Barley leaves | +817 | –83 | –43 | | –75 | | (Torun *et al.*, 2022) |
| ABA | Field | None | Wheat | +30 | –14 | | –54 | | | (Yang *et al.*, 2014) |
| ABA | Greenhouse pot | Drought | Pea | +13 | +33 | | +18 | | | (Latif, 2014) |
| $GA_3$ | Field | Cold | Wheat | | +7.7 | | –11 | | | (Wang *et al.*, 2015) |
| Humic acid seed priming | Growth chamber | None | Rice shoots | | | | +22 | | | (Sheteiwy *et al.*, 2017) |
| Fulvic acid | Growth chamber | Heat | Soybean leaves | | +41 | | | | | (Dinler *et al.*, 2016) |
| Humic acid | Field | Drought | Maize leaves | +19 | –4.5 | | | | | (Chen *et al.*, 2022*c*) |
| Humic acid | Field | None | Mango | +2.9 | | +2.8 | +5.5 | | | (El-Hoseiny *et al.*, 2020) |
| Fulvic acid | Field | None | Maize leaves | +75 | | +66 | +58 | | | (Gao *et al.*, 2022) |
| | | | Maize roots | +50 | | +44 | | | | |
| Chitosan | Greenhouse pot | None | Basil | | +27 | | | –26 | +462 | (Paulert *et al.*, 2021) |
| Algal extracts (*Sargassum tenerrimum*) | Greenhouse pot | None | Tomato leaves | | +63 | –10 | | | +46 | (Khedia *et al.*) |
| Algal extracts (*Ascophyllum nodosum*) | Greenhouse | None | Tomato | +233 | | +35 | +39 | | | (Ali *et al.*, 2022*a*) |
| Algal extracts (*Ascophyllum nodosum*) | Field | None | Hazelnut leaves | +19 | +7.7 | | +6.7 | | | (Cabo *et al.*, 2020) |
| Algal extracts (*Kappaphycus alvarezii*) | Open air pot | Drought | Wheat | | +44 | +125 | | | | (Patel *et al.*, 2018) |
| Algal extracts (*Kappaphycus alvarezii*) | Open air pot | Salt | Wheat | | +20 | +70 | | | | |
| Protein hydrolysates (animal) | Greenhouse pot | Drought | Tomato | +7.7 | | | | | | (Casadesús *et al.*, 2019) |
| *Trichoderma harzianum* | Seedling nursery | None | Melon shoots | +34 | +118 | +29 | | +39 | +77 | (Martínez-Medina *et al.*, 2011) |
| *Glomus mosseae* | Seedling nursery | None | Melon shoots | –54 | –4.6 | | | | –9 | |
| *Trihoderma* and *Glomus mosseae* | Seedling nursery | None | Melon shoots | –8.6 | +27 | +38 | | –11 | +34 | |
| *Glomus intraradices* | Greenhouse pot | Drought | Tomato | | +2.8 | | | | | (Ruiz-Lozano *et al.*, 2016) |
| *Glomus intraradices* | Greenhouse pot | Drought | Lettuce | | +65 | | | | | |
| *Azotobacter vinelandii* | Pot | Salinity | Rice leaves<br>Rice roots | +300<br>+350 | | +532<br>+89 | +200<br>+500 | | | (Sahoo *et al.*, 2014) |
| *Azotobacter vinelandii* | Hydroponic system | Heavy metals | Rice root | +84 | | +155 | +113 | | | (Sahoo *et al.*, 2021) |
| *Bacillus licheniformis* | Greenhouse pot | None | Onion leaves | +29 | | | | | | (Gupta *et al.*, 2021) |
| *Pseudomonas fluorenscence* | Greenhouse pot | None | Onion leaves | +32 | | | | | | |
| *Bacillus subtilis* | Greenhouse pot | None | Cauliflower | –2.8 | +14 | | +29 | | +20 | (Ekinci *et al.*) |
| *Azospirillum brasilense* with N | Field | None | Soybean | +19 | | +45 | +57 | | | (Zahedi and Abbasi, 2015) |

## 5 Summary and the Implications of Regulatory Constraints on The Development and Use of Biostimulants, Microbials and Plant Hormones to Enhance Climate Resilience.

The increased occurrence and severity of extreme weather events because of climate change is among the greatest threats to agricultural productivity globally. Extreme weather events not only reduce crop productivity by compromising crop photosynthesis, metabolism and growth, they also disrupt normal agronomic practices and hence compromise farming efficiency and profitability. The impact of extreme weather events is worsened by the reduced stress tolerance and diminished environmental plasticity of modern cultivars and the reduced resilience of modern cropping systems. Mitigating unpredictable and highly localized extreme weather events will require the development of technologies that can be rapidly and locally implemented to enhance the resilience of the crop to the impending stress. These 'rapid response' technologies will supplement longer term breeding and cropping system strategies that aim to enhance the natural resilience of the crop and the cropping system. Biostimulants represent powerful tools to achieve this goal.

The preceding review has demonstrated that plant tolerance to climate stress is largely mediated through plant sensing mechanisms and the expression and regulation of plant hormone pathways. Plant breeding, agronomic inputs (nutrients and water), plant hormones and biostimulants have all been used to improve plant stress tolerance. In many cases, these benefits are mediated through interactions with natural plant hormone networks. Under both natural and managed conditions, plant hormone levels vary dramatically in response to the local environment and plant response mechanisms (Supplemental Table 1). Biostimulants, microbial amendments and plant growth regulators have shown great promise to strengthen these natural responses, to enhance crop resilience that may have been lost through breeding, selection and modern agronomic practices.

Progress in the utilization of biostimulants, microbials and plant hormones to address the challenges of climate stress is currently constrained by regulatory frameworks that in many parts of the world classify all products that contain or explicitly modulate plant growth regulator pathways as pesticides. The classification of a product as a pesticide is associated with very strict safety requirements that lead to a substantial financial burden on product registration and commercial use, therefore limiting the development and use of such products in agriculture. The requirements imposed on products containing plant hormones or products that modulate plant growth and development derives from US and EU law developed in the 1970's in response to the widespread use of hormones or hormone disruptors as herbicides. This historical ustilization marked the regulatory approach to these products globally. The following statements are illustrative of prevailing global regulations governing the use of pesticides in agriculture (US version shown here):

> US_FIFRA: *With certain exceptions, a pesticide is any substance or mixture of substances intended for preventing, destroying, repelling, or mitigating any pest, or intended for use as a plant regulator.*

The term *'plant regulator'* is further defined as: "any substance or mixture of substances intended, through physiological action, for accelerating or retarding the rate of growth or rate of maturation or altering the behavior of plants." It is specifically stated in EPA rulings that all known plant hormones are considered pesticides and that products that contain known plant hormones or that explicitly claim to act through the modification of these pathways would be deemed pesticidal.

The breadth of this regulatory judgement is problematic as it implies that all products that alter plant growth are *pesticides* unless explicitly exempted through subsequent regulatory rulings. Interpreted literally, this definition would imply that plant breeding or the use of plant nutrients to improve plant growth and development could be regarded as pesticides. Recognizing the undue burden this definition poses on normal agricultural practices such as fertilization, irrigation and soil amendments, the US EPA has created a positive list of exempt product classes specifying that " a product of any of the following types, intended only to aid the growth of desirable plants, is not a "plant regulator" under section 2(v) of FIFRA, and therefore is not a pesticide. (FIFRA Amendment 40 CFR § 152.6). In 2024 the US proposed, through introduction of a bill to the US House of Representatives, that biostimulants would obtain a similar exemption under FIFRA. In the EU, biostimulants are currently exempt from pesticide Regulation (EC) 1107/2009 according to its Article 2(b) "influencing the life processes of plants, such as substances influencing their growth, other than as a nutrient or a plant biostimulant", and they are classified as one category of fertilizing products (2019/1009, 2019).

While the proposed US biostimulant exemption and the existing EU biostimulant category appear to offer pathways for the use of these products to alleviate plant stress, the conflict with existing plant growth regulator definitions, and the evidence provided here that crop resilience is mediated through changes in plant growth and development, involving changes in endogenous hormone metabolism, represents a substantial scientific and marketing impediment to the full rational development and implementation of these promising tools. The inability of manufacturers of biostimulants, plant hormones or plant growth regulators to use these products to beneficially target plant regulatory networks, or beneficially modulate plant hormone pathways, or publicly claim such effects on labels or marketing materials, severely constrains progress. The implication that manufacturers of biostimulants must avoid the suggestion that enhanced plant stress tolerance is mediated by plant regulatory pathways is antithetical to the underlying science. It is also inconsistent with natural resilience mechanisms where substantial fluctuations in internal plant regulatory pathways and hormone concentrations is the means by which climate resilience is attained. Changes in plant hormone levels occur naturally in all plants as a response to stress or the application of nutrients or water. These natural responses often result in changes in plant hormone concentrations and plant growth effects that do not exceed those seen from the application of biostimulants or plant hormones when used at rates beneficial for the applied plant.

## 6 Conclusions and future prospectives

Solutions will be needed for both the long-term effects of climate change, and the more immediate and critical impacts of more frequent extreme weather events that increasingly cause profound agronomic disruption. These challenges can be addressed by methods including targeted breeding, increased resource inputs and the application of plant biostimulants, plant hormones and other strategies that aim to enhance crop resilience and adaptation. Targeted breeding can provide robust solutions for long term climate threats, but will require an extended investment of time and money, and must be replicated for each discrete cropping system. This process is inherently too slow to address the unpredictable and highly local extreme climate events.

There is a clear need for more readily implementable solutions that can help mitigate the impact of extreme weather conditions such as frost, heat, drought, flooding, etc. Biostimulants, plant hormones and microbial inoculants have tremendous potential to provide tools that are more flexible and rapidly implementable alternatives to breeding. Coupling the application of biostimulants or plant hormones with weather prediction and just-in-time precision application has the potential to reduce the negative effects of abiotic stress on crop production while offering a highly tailored solution to local challenges.

The use of biostimulants, hormones or microbials and other products to address the threat of climate change, extreme weather and abiotic stress is novel and as such does not have a suitable enabling regulatory framework. An opportunity exists to utilize best available knowledge of plant growth principles to develop regulatory guidelines that will allow the development of these products to safely and effectively maximize crop productivity while safeguarding the environment.

The lack of globally accepted regulatory standards for the design and use of biostimulants has stifled their development, hindered their application and compromised grower acceptance. The current scientifically unsound and outdated legislative paradigm specifying 'that any product acting directly upon plant growth processes' is de-facto 'pesticidal' represents the greatest barrier to the development and use of these products in agriculture. Given that most plant hormones when present at concentrations found in nature have been designated as safe, with no viable human toxicity nor environmental persistence, there appears to be no justification for limiting their use in agriculture when 'used for the benefit of the target crop at naturally occurring concentrations'. In addition, these products act in harmony with physiological processes, minimizing ecological impact while maximizing their agronomic efficacy. Likewise, the use of biostimulants or microbial products or inorganic substances that are themselves deem safe, should not be considered pesticidal merely because they directly alter natural plant growth regulatory pathways.

There is an immediate need to differentiate between products of negligible toxicity that are used solely for the benefit of the targeted crop (including plant growth regulators and biostimulants) and those products that are used solely to constrain or impede the growth of non-beneficial plants (herbicides). Classification of products with dual use components according to the rate of use or intent of use of the product is easily implemented, with many examples of such frameworks already in place.

Regulatory clarity is necessary to foster innovation and enable the deployment of biostimulants to help reduce the negative impacts of climate change and extreme weather on crop production. These are powerful tools with tremendous potential to enhance crop productivity in the face of stress and hence improve the efficiency of use of inputs such as water, nutrients and pesticides.

**Acknowledgment**

The BPIA (Biological Products Industry Alliance, US) provided financial assistance to GF and JL at the request of PB. BPIA did not influence the content of this review and did not contribute to writing or interpretation.

**Reference**

**2019/1009 R (EU)**. 2019. *Regulation (EU) 2019/1009*.

**Abouelsaad I, Renault S**. 2018. Enhanced oxidative stress in the jasmonic acid-deficient tomato mutant def-1 exposed to NaCl stress. Journal of plant physiology **226**, 136–144.

**Adil Aydin**. 2012. Humic acid application alleviate salinity stress of bean (Phaseolus vulgaris L.) plants decreasing membrane leakage. AFRICAN JOURNAL OF AGRICULTURAL RESEEARCH **7**.

**Ahmad RM, Cheng C, Sheng J, Wang W, Ren H, Aslam M, Yan Y**. 2019. Interruption of Jasmonic Acid Biosynthesis Causes Differential Responses in the Roots and Shoots of Maize Seedlings against Salt Stress. International Journal of Molecular Sciences 2019, Vol. 20, Page 6202 **20**, 6202.

**Ahmed M, Ullah H, Piromsri K, Tisarum R, Cha-um S, Datta A**. 2022. Effects of an Ascophyllum nodosum seaweed extract application dose and method on growth, fruit yield, quality, and water productivity of tomato under water-deficit stress. South African Journal of Botany **151**, 95–107.

**Ali S, Moon YS, Hamayun M, Khan MA, Bibi K, Lee IJ**. 2022. Pragmatic role of microbial plant biostimulants in abiotic stress relief in crop plants. Taylor and Francis Ltd.

**Ali O, Ramsubhag A, Daniram Benn Jr. Ramnarine S, Jayaraman J**. 2022*a*. Transcriptomic changes induced by applications of a commercial extract of Ascophyllum nodosum on tomato plants. Scientific Reports **12**.

**Ali AH, Said EM, Abdelgawad ZA**. 2022*b*. The role of seaweed extract on improvement drought tolerance of wheat revealed by osmoprotectants and DNA (cpDNA) markers. Revista Brasileira de Botanica **45**, 857–867.

**Alvarez ME, Savouré A, Szabados L**. 2022. Proline metabolism as regulatory hub. Trends in Plant Science **27**, 39–55.

**Alves LR, Monteiro CC, Carvalho RF, Ribeiro PC, Tezotto T, Azevedo RA, Gratão PL**. 2017. Cadmium stress related to root-to-shoot communication depends on ethylene and auxin in tomato plants. Environmental and Experimental Botany **134**, 102–115.

**Ambreetha S, Chinnadurai C, Marimuthu P, Balachandar D**. 2018. Plant-associated Bacillus modulates the expression of auxin-responsive genes of rice and modifies the root architecture. Rhizosphere **5**, 57–66.

**Arnold PA, Kruuk LEB, Nicotra AB**. 2019. How to analyse plant phenotypic plasticity in response to a changing climate. New Phytologist **222**, 1235–1241.

**Arslan E, Agar G, Aydin M**. Humic Acid as a Biostimulant in Improving Drought Tolerance in Wheat: the Expression Patterns of Drought-Related Genes. doi: 10.1007/s11105-020-01266-3/Published.

**Bado S, Forster BP, Nielen S, Ali AM, Lagoda PJL, Till BJ, Laimer M**. 2015. Plant mutation breeding: current progress and future assessment. Plant Breeding Reviews: Volume 39, 23–88.

**Baena-González E, Hanson J**. 2017. Shaping plant development through the SnRK1–TOR metabolic regulators. Current Opinion in Plant Biology **35**, 152–157.

**Barnawal D, Bharti N, Pandey SS, Pandey A, Chanotiya CS, Kalra A**. 2017. Plant growth-promoting rhizobacteria enhance wheat salt and drought stress tolerance by altering endogenous phytohormone levels and TaCTR1/TaDREB2 expression. Physiologia Plantarum **161**, 502–514.

**Barra Caracciolo A, Terenzi V**. 2021. Rhizosphere microbial communities and heavy metals. MDPI AG.

**Bi CL, Wen XJ, Zhang XY, Liu X**. 2010. Cloning and characterization of a putative CTR1 gene from wheat. Agricultural Sciences in China **9**, 1241–1250.

**Bilal S, Shahzad R, Imran M, Jan R, Kim KM, Lee IJ**. 2020. Synergistic association of endophytic fungi enhances Glycine max L. resilience to combined abiotic stresses: Heavy metals, high temperature and drought stress. Industrial Crops and Products **143**.

**Bouteraa MT, Mishra A, Romdhane W Ben, Hsouna A Ben, Siddique KHM, Saad R Ben**. 2022. Bio-Stimulating Effect of Natural Polysaccharides from Lobularia maritima on Durum Wheat Seedlings: Improved Plant Growth, Salt Stress Tolerance by Modulating Biochemical Responses and Ion Homeostasis. Plants **11**.

**Bouzroud S, Gasparini K, Hu G, Barbosa MAM, Rosa BL, Fahr M, Bendaou N, Bouzayen M, Zsögön A, Smouni A**. 2020. Down regulation and loss of auxin response factor 4 function using CRISPR/Cas9 alters plant growth, stomatal function and improves tomato tolerance to salinity and osmotic stress. Genes **11**, 272.

**Brecht JK**. 2019. Ethylene Technology. Postharvest Technology of Perishable Horticultural Commodities, 481–497.

**Brooker R, Brown LK, George TS, Pakeman RJ, Palmer S, Ramsay L, Schöb C, Schurch N, Wilkinson MJ**. 2022. Active and adaptive plasticity in a changing climate. Trends in Plant Science **27**, 717–728.

**Burney J, McIntosh C, Lopez-Videla B, Samphantharak K, Maia AG**. 2024. Empirical modeling of agricultural climate risk. Proceedings of the National Academy of Sciences of the United States of America **121**, e2215677121.

**Busi R, Goggin DE, Heap IM, *et al.*** 2018. Weed resistance to synthetic auxin herbicides. Pest Management Science **74**, 2265–2276.

**Cabo S, Morais MC, Aires A, Carvalho R, Pascual-Seva N, Silva AP, Gonçalves B**. 2020. Kaolin and seaweed-based extracts can be used as middle and long-term strategy to mitigate negative effects of climate change in physiological performance of hazelnut tree. Journal of Agronomy and Crop Science **206**, 28–42.

**Canellas LP, Olivares FL, Aguiar NO, Jones DL, Nebbioso A, Mazzei P, Piccolo A**. 2015. Humic and fulvic acids as biostimulants in horticulture. Elsevier.

**Canellas LP, da Silva RM, Busato JG, Olivares FL**. 2024. Humic substances and plant abiotic stress adaptation. Springer Science and Business Media Deutschland GmbH.

**Carillo P, Ciarmiello LF, Woodrow P, Corrado G, Chiaiese P, Rouphael Y**. 2020. Enhancing sustainability by improving plant salt tolerance through macro-and micro-algal biostimulants. MDPI AG.

**Casadesús A, Pérez-Llorca M, Munné-Bosch S, Polo J**. 2020. An Enzymatically Hydrolyzed Animal Protein-Based Biostimulant (Pepton) Increases Salicylic Acid and Promotes Growth of Tomato Roots Under Temperature and Nutrient Stress. Frontiers in Plant Science **11**.

**Casadesús A, Polo J, Munné-Bosch S**. 2019. Hormonal effects of an enzymatically hydrolyzed animal protein-based biostimulant (pepton) in water-stressed tomato plants. Frontiers in Plant Science **10**.

**Ceccarelli AV, Miras-Moreno B, Buffagni V, Senizza B, Pii Y, Cardarelli M, Rouphael Y, Colla G, Lucini L**. 2021. Foliar application of different vegetal-derived protein hydrolysates distinctively modulates tomato root development and metabolism. Plants **10**, 1–15.

**Chanda M joan, Merghoub N, EL Arroussi H**. 2019. Microalgae polysaccharides: the new sustainable bioactive products for the development of plant bio-stimulants? Springer Netherlands.

**Chang YN, Zhu C, Jiang J, Zhang H, Zhu JK, Duan CG**. 2020. Epigenetic regulation in plant abiotic stress responses. Journal of Integrative Plant Biology **62**, 563–580.

**Chen Y, Cothren JT, Chen D, Ibrahim AMH, Lombardini L**. 2014. Effect of 1-MCP on cotton plants under abiotic stress caused by ethephon. American Journal of Plant Sciences **5**, 3005.

**Chen Y, Jin M, Wu CY, Bao JP**. 2022*a*. Effects of Plant Growth Regulators on the Endogenous Hormone Content of Calyx Development in 'Korla' Fragrant Pear. HortScience **57**, 497–503.

**Chen A, Koehler AN**. 2020. Transcription Factor Inhibition: Lessons Learned and Emerging Targets. Trends in Molecular Medicine **26**, 508–518.

**Chen Q, Qu Z, Ma G, Wang W, Dai J, Zhang M, Wei Z, Liu Z**. 2022*b*. Humic acid modulates growth, photosynthesis, hormone and osmolytes system of maize under drought conditions. Agricultural Water Management **263**.

**Chen Q, Qu Z, Ma G, Wang W, Dai J, Zhang M, Wei Z, Liu Z**. 2022*c*. Humic acid modulates growth, photosynthesis, hormone and osmolytes system of maize under drought conditions. Agricultural Water Management **263**.

**Chen Z, Zhou W, Guo X, Ling S, Li W, Wang X, Yao J**. 2024. Heat Stress Responsive Aux/IAA Protein, OsIAA29 Regulates Grain Filling Through OsARF17 Mediated Auxin Signaling Pathway. Rice **17**, 1–14.

**Chesterfield RJ, Vickers CE, Beveridge CA**. 2020. Translation of Strigolactones from Plant Hormone to Agriculture: Achievements, Future Perspectives, and Challenges. Trends in Plant Science **25**, 1087–1106.

**Colla G, Hoagland L, Ruzzi M, Cardarelli M, Bonini P, Canaguier R, Rouphael Y**. 2017. Biostimulant action of protein hydrolysates: Unraveling their effects on plant physiology and microbiome. Frontiers Media S.A.

**Colla G, Nardi S, Cardarelli M, Ertani A, Lucini L, Canaguier R, Rouphael Y**. 2015. Protein hydrolysates as biostimulants in horticulture. Elsevier.

**Colla G, Rouphael Y, Canaguier R, Svecova E, Cardarelli M**. 2014. Biostimulant action of a plant-derived protein hydrolysate produced through enzymatic hydrolysis. Frontiers in Plant Science **5**.

**Colla G, Rouphael Y, Di Mattia E, El-Nakhel C, Cardarelli M**. 2015. Co-inoculation of Glomus intraradices and Trichoderma atroviride acts as a biostimulant to promote growth, yield and nutrient uptake of vegetable crops. Journal of the Science of Food and Agriculture **95**, 1706–1715.

**Deolu-Ajayi AO, van der Meer IM, van der Werf A, Karlova R**. 2022. The power of seaweeds as plant biostimulants to boost crop production under abiotic stress. John Wiley and Sons Inc.

**Depaepe T, Van D, Straeten D, Depaepe T, Van Der Straeten D**. 2020. Tools of the Ethylene Trade: A Chemical Kit to Influence Ethylene Responses in Plants and Its Use in Agriculture. Small Methods **4**, 1900267.

**Dietz KJ, Vogelsang L**. 2024. A general concept of quantitative abiotic stress sensing. Trends in Plant Science **29**, 319–328.

**Dinler BS, Gunduzer E, Tekinay T**. 2016. Pre-treatment of fulvic acid plays a stimulant role in protection of soybean (Glycine max L.) leaves against heat and salt stress. Acta Biologica Cracoviensia Series Botanica **58**, 29–41.

**Djanaguiraman M, Prasad PV V, Al-Khatib K**. 2011. Ethylene perception inhibitor 1-MCP decreases oxidative damage of leaves through enhanced antioxidant defense mechanisms in soybean plants grown under high temperature stress. Environmental and experimental botany **71**, 215–223.

**Du H, Liu H, Xiong L**. 2013. Endogenous auxin and jasmonic acid levels are differentially modulated by abiotic stresses in rice. Frontiers in Plant Science **4**.

**Du H, Wu N, Cui F, You L, Li X, Xiong L**. 2014. A homolog of ETHYLENE OVERPRODUCER, OsETOL1, differentially modulates drought and submergence tolerance in rice. The Plant Journal **78**, 834–849.

**Du H, Wu N, Fu J, Wang S, Li X, Xiao J, Xiong L**. 2012. A GH3 family member, OsGH3-2, modulates auxin and abscisic acid levels and differentially affects drought and cold tolerance in rice. Journal of experimental botany **63**, 6467–6480.

**Dubois M, Van den Broeck L, Inzé D**. 2018. The pivotal role of ethylene in plant growth. Trends in plant science **23**, 311–323.

**Ekinci M, Turan M, Yildirim E**. *Effect of plant growth promoting rhizobacteria on growth, nutrient, organic acid, amino acid and hormone content of cauliflower (Brassica oleracea L. var. botrytis) transplants*.

**Elansary HO, Skalicka-Woźniak K, King IW**. 2016. Enhancing stress growth traits as well as phytochemical and antioxidant contents of Spiraea and Pittosporum under seaweed extract treatments. Plant Physiology and Biochemistry **105**, 310–320.

**El-Hoseiny HM, Helaly MN, Elsheery NI, Alam-Eldein SM**. 2020. Humic acid and boron to minimize the incidence of alternate bearing and improve the productivity and fruit quality of mango trees. HortScience **55**, 1026–1037.

**Enebe MC, Babalola OO**. 2018. The influence of plant growth-promoting rhizobacteria in plant tolerance to abiotic stress: a survival strategy. Springer Verlag.

**EPA**. 2024. *Draft Guidance for Plant Regulators and Claims, Including Plant Biostimulants | US EPA*.

**Etehadnia M, Waterer DR, Tanino KK**. 2008. The Method of ABA Application Affects Salt Stress Responses in Resistant and Sensitive Potato Lines. JOURNAL OF PLANT GROWTH REGULATION **27**, 331–341.

**Etesami H**. 2018. Bacterial mediated alleviation of heavy metal stress and decreased accumulation of metals in plant tissues: Mechanisms and future prospects. Academic Press.

**Fàbregas N, Fernie AR**. 2021. The interface of central metabolism with hormone signaling in plants. Current Biology **31**, R1535–R1548.

**Fitzpatrick CR, Copeland J, Wang PW, Guttman DS, Kotanen PM, Johnson MTJ**. Assembly and ecological function of the root microbiome across angiosperm plant species. doi: 10.5061/dryad.5p414.

**Francesca S, Cirillo V, Raimondi G, Maggio A, Barone A, Rigano MM**. 2021. A Novel Protein Hydrolysate-Based Biostimulant Improves Tomato Performances under Drought Stress. Plants **10**.

**Francesca S, Najai S, Zhou R, Decros G, Cassan C, Delmas F, Ottosen CO, Barone A, Rigano MM**. 2022. Phenotyping to dissect the biostimulant action of a protein hydrolysate in tomato plants under combined abiotic stress. Plant Physiology and Biochemistry **179**, 32–43.

**Franzoni G, Cocetta G, Prinsi B, Ferrante A, Espen L**. 2022. Biostimulants on Crops: Their Impact under Abiotic Stress Conditions. MDPI.

**Furio RN, Salazar SM, Mariotti-Martínez JA, Martínez-Zamora GM, Coll Y, Díaz-Ricci JC**. 2022. Brassinosteroid Applications Enhance the Tolerance to Abiotic Stresses, Production and Quality of Strawberry Fruits. **8**, 572.

**Gallusci P, Agius DR, Moschou PN, Dobránszki J, Kaiserli E, Martinelli F**. 2023. Deep inside the epigenetic memories of stressed plants. Trends in Plant Science **28**, 142–153.

**Gao F, Li Z, Du Y, Duan J, Zhang T, Wei Z, Guo L, Gong W, Liu Z, Zhang M**. 2022. The Combined Application of Urea and Fulvic Acid Solution Improved Maize Carbon and Nitrogen Metabolism. Agronomy **12**.

**Gao R, Yu Q, Shen Y, Chu Q, Chen G, Fen S, Yang M, Yuan L, McClements DJ, Sun Q**. 2021. Production, bioactive properties, and potential applications of fish protein hydrolysates: Developments and challenges. Elsevier Ltd.

**Gashaw A, Theerawitaya C, Samphumphuang T, Cha-um S, Supaibulwatana K**. 2014. CPPU elevates photosynthetic abilities, growth performances and yield traits in salt stressed rice (Oryza sativa L. spp. indica) via free proline and sugar accumulation. Pesticide Biochemistry and Physiology **108**, 27–33.

**Gemrotová M, Kulkarni MG, Stirk WA, Strnad M, Van Staden J, Spíchal L**. 2013. Seedlings of medicinal plants treated with either a cytokinin antagonist (PI-55) or an inhibitor of cytokinin degradation (INCYDE) are protected against the negative effects of cadmium. Plant Growth Regulation **71**, 137–145.

**Glick BR**. 2014. Bacteria with ACC deaminase can promote plant growth and help to feed the world. Microbiological Research **169**, 30–39.

**Grillo S, Leone A, Xu Y, Tucci M, Francione R, Hasegawa PM, Monti L, Bressan RA**. 1995. Control of osmotin gene expression by ABA and osmotic stress in vegetative tissues of wild-type and ABA-deficient mutants of tomato. Physiologia Plantarum **93**, 498–504.

**Gujjar RS, Banyen P, Chuekong W, Worakan P, Roytrakul S, Supaibulwatana K**. 2020. A Synthetic Cytokinin Improves Photosynthesis in Rice under Drought Stress by Modulating the Abundance of Proteins Related to Stomatal Conductance, Chlorophyll Contents, and Rubisco Activity. Plants 2020, Vol. 9, Page 1106 **9**, 1106.

**Gupta S, Stirk WA, Plačková L, Kulkarni MG, Doležal K, Van Staden J**. 2021. Interactive effects of plant growth-promoting rhizobacteria and a seaweed extract on the growth and physiology of Allium cepa L. (onion). Journal of Plant Physiology **262**.

**Gurmani AR, Bano A, Khan SU, Din J, Zhang JL**. 2011. Alleviation of salt stress by seed treatment with abscisic acid (ABA), 6-benzylaminopurine (BA) and chlormequat chloride (CCC) optimizes ion and organic matter accumulation and increases yield of rice ('Oryza sativa'L.). Australian Journal of Crop Science **5**, 1278–1285.

**Hamayun M, Khan SA, Khan AL, Shin JH, Ahmad B, Shin DH, Lee IJ**. 2010. Exogenous gibberellic acid reprograms soybean to higher growth and salt stress tolerance. Journal of Agricultural and Food Chemistry **58**, 7226–7232.

**Han YJ, Kim YS, Hwang OJ, Roh J, Ganguly K, Kim SK, Hwang I, Kim J Il**. 2017. Overexpression of Arabidopsis thaliana brassinosteroid-related acyltransferase 1 gene induces brassinosteroid-deficient phenotypes in creeping bentgrass. PLOS ONE **12**, e0187378.

**Hasegawa T, Sakurai G, Fujimori S, Takahashi K, Hijioka Y, Masui T**. 2021. Extreme climate events increase risk of global food insecurity and adaptation needs. Nature Food 2021 2:8 **2**, 587–595.

**He N, Li Y, Liu C, *et al.*** 2020. Plant Trait Networks: Improved Resolution of the Dimensionality of Adaptation. Trends in Ecology & Evolution **35**, 908–918.

**He F, Sheng M, Tang M**. 2017. Effects of Rhizophagus irregularis on photosynthesis and antioxidative enzymatic system in Robinia pseudoacacia L. under drought stress. Frontiers in Plant Science **8**, 183.

**He Z, Webster S, He SY**. 2022. Growth–defense trade-offs in plants. Current Biology **32**, R634–R639.

**Hernández-Herrera RM, Sánchez-Hernández CV, Palmeros-Suárez PA, Ocampo-Alvarez H, Santacruz-Ruvalcaba F, Meza-Canales ID, Becerril-Espinosa A**. 2022. Seaweed Extract Improves Growth and Productivity of Tomato Plants under Salinity Stress. Agronomy **12**.

**Ho S-L, Chao Y-C, Tong W-F, Yu S-M**. 2001. *Sugar Coordinately and Differentially Regulates Growth-and Stress-Related Gene Expression via a Complex Signal Transduction Network and Multiple Control Mechanisms 1*.

**Huang H, Liu B, Liu L, Song S**. 2017. Jasmonate action in plant growth and development. Journal of Experimental Botany **68**, 1349–1359.

**Huang Z, Zhang Z, Zhang X, Zhang H, Huang D, Huang R**. 2004. Tomato TERF1 modulates ethylene response and enhances osmotic stress tolerance by activating expression of downstream genes. Febs Letters **573**, 110–116.

**Hussain S, Bai Z, Huang J, Cao X, Zhu L, Zhu C, Khaskheli MA, Zhong C, Jin Q, Zhang J**. 2019. 1-Methylcyclopropene modulates physiological, biochemical, and antioxidant responses of rice to different salt stress levels. Frontiers in plant science **10**, 124.

**Illescas M, Pedrero-Méndez A, Pitorini-Bovolini M, Hermosa R, Monte E**. 2021. Phytohormone production profiles in Trichoderma species and their relationship to wheat plant responses to water stress. Pathogens **10**, 991.

**Iqbal M, Ashraf M**. 2013. Gibberellic acid mediated induction of salt tolerance in wheat plants: Growth, ionic partitioning, photosynthesis, yield and hormonal homeostasis. Environmental and Experimental Botany **86**, 76–85.

**Iqbal N, Umar S, Per TS, Khan NA**. 2017. Ethephon increases photosynthetic-nitrogen use efficiency, proline and antioxidant metabolism to alleviate decrease in photosynthesis under salinity stress in mustard. Plant Signaling & Behavior **12**.

**ISO**. ISO 8157:2022 - Fertilizers, soil conditioners and beneficial substances — Vocabulary. https://www.iso.org/standard/80949.html. Accessed October 2024.

**Jägermeyr J, Müller C, Ruane AC, *et al.*** 2021. Climate impacts on global agriculture emerge earlier in new generation of climate and crop models. Nature Food 2021 2:11 **2**, 873–885.

**Jan SS, Khan NA, Asaf S, Shahzad R, Lubna, Imran M, Bilal S, Lee IJ, Al-Harrasi A**. 2024. Consortium of Endophytic Bacillus australimaris CK11 and Staphylococcus epidermidis CK9 from Commiphora gileadensis Mediates Tomato Resilience to Combined Salinity, Heat, and Drought Stresses. Journal of Plant Growth Regulation doi: 10.1007/s00344-024-11394-9.

**Jankowicz-Cieslak J, Till BJ**. 2015. Forward and reverse genetics in crop breeding. Advances in plant breeding strategies: breeding, biotechnology and molecular tools, 215–240.

**Jayakodi M, Padmarasu S, Haberer G, Bonthala VS, Gundlach H, Monat C, Lux T, Kamal N, Lang D, Himmelbach A**. 2020. The barley pan-genome reveals the hidden legacy of mutation breeding. Nature **588**, 284–289.

**Jha CK, Sharma P, Shukla A, Parmar P, Patel R, Goswami D, Saraf M**. 2021. Microbial enzyme, 1-aminocyclopropane-1-carboxylic acid (ACC) deaminase: An elixir for plant under stress. Academic Press.

**Jiang W, Wang X, Wang Y, Du Y, Zhang S, Zhou H, Feng N, Zheng D, Ma G, Zhao L**. 2024. S-ABA Enhances Rice Salt Tolerance by Regulating Na+/K+ Balance and Hormone Homeostasis. Metabolites **14**, 181.

**Jiang M, Xu F, Peng M, Huang F, Meng F**. 2016. Methyl jasmonate regulated diploid and tetraploid black locust (Robinia pseudoacacia L.) tolerance to salt stress. Acta Physiologiae Plantarum **38**, 1–13.

**Jin P, Zhu H, Wang J, Chen J, Wang X, Zheng Y**. 2013. Effect of methyl jasmonate on energy metabolism in peach fruit during chilling stress. Journal of the Science of Food and Agriculture **93**, 1827–1832.

**Jini D, Joseph B**. 2017. Physiological mechanism of salicylic acid for alleviation of salt stress in rice. Rice Science **24**, 97–108.

**Jones AM**. 2016. A new look at stress: abscisic acid patterns and dynamics at high-resolution. New Phytologist **210**, 38–44.

**Joshi PM, Juwarkar AA**. 2009. In vivo studies to elucidate the role of extracellular polymeric substances from Azotobacter in immobilization of heavy metals. Environmental Science and Technology **43**, 5884–5889.

**Kamínek M**. 2015. Tracking the Story of Cytokinin Research. Journal of Plant Growth Regulation 2015 34:4 **34**, 723–739.

**Kang SM, Radhakrishnan R, Khan AL, Kim MJ, Park JM, Kim BR, Shin DH, Lee IJ**. 2014. Gibberellin secreting rhizobacterium, Pseudomonas putida H-2-3 modulates the hormonal and stress physiology of soybean to improve the plant growth under saline and drought conditions. Plant Physiology and Biochemistry **84**, 115–124.

**Kang S-M, Shahzad R, Bilal S, Khan AL, Park Y-G, Lee K-E, Asaf S, Khan MA, Lee I-J**. 2019. Indole-3-acetic-acid and ACC deaminase producing Leclercia adecarboxylata MO1 improves Solanum lycopersicum L. growth and salinity stress tolerance by endogenous secondary metabolites regulation. BMC microbiology **19**, 1–14.

**Kasim W**. 2017. Alleviation of Drought Stress in Vicia faba by Seed Priming with Ascorbic Acid or Extracts of Garlic and Carrot. Egyptian Journal of Botany **0**, 45–59.

**Kavita B, Shukla S, Naresh Kumar G, Archana G**. 2008. Amelioration of phytotoxic effects of Cd on mung bean seedlings by gluconic acid secreting rhizobacterium Enterobacter asburiae

PSI3 and implication of role of organic acid. World Journal of Microbiology and Biotechnology **24**, 2965–2972.

**Keswani C, Singh SP, Cueto L, García-Estrada C, Mezaache-Aichour S, Glare TR, Borriss R, Singh SP, Blázquez MA, Sansinenea E**. 2020. Auxins of microbial origin and their use in agriculture. Applied Microbiology and Biotechnology **104**, 8549–8565.

**Khan MS, Akther T, Ali DM, Hemalatha S**. 2019. An investigation on the role of salicylic acid alleviate the saline stress in rice crop (Oryza sativa (L)). Biocatalysis and Agricultural Biotechnology **18**, 101027.

**Khan MA, Asaf S, Khan AL, Jan R, Kang S-M, Kim K-M, Lee I-J**. 2020. Extending thermotolerance to tomato seedlings by inoculation with SA1 isolate of Bacillus cereus and comparison with exogenous humic acid application. PLoS One **15**, e0232228.

**Khan I, Awan SA, Ikram R, Rizwan M, Akhtar N, Yasmin H, Sayyed RZ, Ali S, Ilyas N**. 2021*a*. Effects of 24-epibrassinolide on plant growth, antioxidants defense system, and endogenous hormones in two wheat varieties under drought stress. Physiologia plantarum **172**, 696–706.

**Khan Z, Gul H, Rauf M, *et al.*** 2022. Sargassum wightii Aqueous Extract Improved Salt Stress Tolerance in Abelmoschus esculentus by Mediating Metabolic and Ionic Rebalance. Frontiers in Marine Science **9**.

**Khan MA, Hamayun M, Asaf S, Khan M, Yun BW, Kang SM, Lee IJ**. 2021*b*. Rhizospheric Bacillus spp. Rescues Plant Growth Under Salinity Stress via Regulating Gene Expression, Endogenous Hormones, and Antioxidant System of Oryza sativa L. Frontiers in Plant Science **12**, 665590.

**Khedia J, Dangariya M, Nakum AK, Agarwal P, Panda A, Kumar Parida A, Gangapur DR, Meena R, Agarwal PK**. Sargassum seaweed extract enhances Macrophomina phaseolina resistance in tomato by regulating phytohormones and antioxidative activity. doi: 10.1007/s10811-020-02263-5/Published.

**Khoury CK, Brush S, Costich DE, *et al.*** 2022. Crop genetic erosion: understanding and responding to loss of crop diversity. New Phytologist **233**, 84–118.

**Kim EH, Kim YS, Park S-H, Koo YJ, Choi Y Do, Chung Y-Y, Lee I-J, Kim J-K**. 2009. Methyl jasmonate reduces grain yield by mediating stress signals to alter spikelet development in rice. Plant physiology **149**, 1751–1760.

**Krishnan S, Merewitz EB**. 2015. Drought Stress and Trinexapac-ethyl Modify Phytohormone Content Within Kentucky Bluegrass Leaves. Journal of Plant Growth Regulation **34**, 1–12.

**Kuromori T, Seo M, Shinozaki K**. 2018. ABA Transport and Plant Water Stress Responses. Trends in Plant Science **23**, 513–522.

**Landis JB, Guercio AM, Brown KE, Fiscus CJ, Morrell PL, Koenig D**. 2024. Natural selection drives emergent genetic homogeneity in a century-scale experiment with barley. Science **385**, eadl0038.

**Latif HH**. 2014. *PHYSIOLOGICAL RESPONSES OF PISUM SATIVUM PLANT TO EXOGENOUS ABA APPLICATION UNDER DROUGHT CONDITIONS*.

**Latif Khan A, Hamayun M, Kang S-M, Kim Y-H, Jung H-Y, Lee J-H, Lee I-J**. 2012. *Endophytic fungal association via gibberellins and indole acetic acid can improve plant growth under abiotic stress: an example of Paecilomyces formosus LHL10*.

**Latif HH, Mohamed HI**. 2016*a*. Exogenous applications of moringa leaf extract effect on retrotransposon, ultrastructural and biochemical contents of common bean plants under environmental stresses. South African Journal of Botany **106**, 221–231.

**Latif HH, Mohamed HI**. 2016*b*. Exogenous applications of moringa leaf extract effect on retrotransposon, ultrastructural and biochemical contents of common bean plants under environmental stresses. South African Journal of Botany **106**, 221–231.

**Lee SK, Sohn EY, Hamayun M, Yoon JY, Lee IJ**. 2010. Effect of silicon on growth and salinity stress of soybean plant grown under hydroponic system. Agroforestry Systems **80**, 333–340.

**Li L, Chen X**. 2023. Auxin regulation on crop: from mechanisms to opportunities in soybean breeding. Molecular Breeding 2023 43:3 **43**, 1–26.

**Li JH, Feng NJ, Zheng DF, Du X Le, Wu JS, Wang X**. 2024. Regulation of seed soaking with indole-3-butyric acid potassium salt (IBA-K) on rapeseed (Brassica napus L.) seedlings under NaCl stress. BMC plant biology **24**, 904.

**Li J, Van Gerrewey T, Geelen D**. 2022. A Meta-Analysis of Biostimulant Yield Effectiveness in Field Trials. Frontiers Media S.A.

**Li L, Gu W, Li J, Li C, Xie T, Qu D, Meng Y, Li C, Wei S**. 2018. Exogenously applied spermidine alleviates photosynthetic inhibition under drought stress in maize (Zea mays L.) seedlings associated with changes in endogenous polyamines and phytohormones. Plant Physiology and Biochemistry **129**, 35–55.

**Li Q, Wang G, Guan C, Yang D, Wang Y, Zhang Y, Ji J, Jin C, An T**. 2019. Overexpression of LcSABP, an orthologous gene for salicylic acid binding protein 2, enhances drought stress tolerance in transgenic tobacco. Frontiers in Plant Science **10**, 435271.

**Lichtfouse E**. 2021. *Sustainable Agriculture Reviews*. Springer Nature.

**Liu R, Liang G, Gong J, Wang J, Zhang Y, Hao Z, Li G**. 2023. A Potential ABA Analog to Increase Drought Tolerance in Arabidopsis thaliana. International Journal of Molecular Sciences **24**, 8783.

**Liu C-H, Siew W, Hung Y-T, Jiang Y-T, Huang C-H**. 2021. 1-Aminocyclopropane-1-carboxylate (ACC) deaminase gene in Pseudomonas azotoformans is associated with the amelioration of salinity stress in tomato. Journal of Agricultural and Food Chemistry **69**, 913–921.

**Liu L, Sun Y, Di P, Cui Y, Meng Q, Wu X, Chen Y, Yuan J**. 2022. Overexpression of a Zea mays Brassinosteroid-Signaling Kinase Gene ZmBSK1 Confers Salt Stress Tolerance in Maize. Frontiers in Plant Science **13**, 894710.

**Lubovská Z, Dobrá J, Štorchová H, Wilhelmová N, Vanková R**. 2014. Cytokinin oxidase/dehydrogenase overexpression modifies antioxidant defense against heat, drought and their combination in Nicotiana tabacum plants. Journal of Plant Physiology **171**, 1625–1633.

**Luqman M, Shahbaz M, Maqsood MF, Farhat F, Zulfiqar U, Siddiqui MH, Masood A, Aqeel M, Haider FU**. 2023. Effect of strigolactone on growth, photosynthetic efficiency, antioxidant activity, and osmolytes accumulation in different maize (Zea mays L.) hybrids grown under drought stress. Plant Signaling and Behavior **18**.

**Ma C, Bian C, Liu W, Sun Z, Xi X, Guo D, Liu X, Tian Y, Wang C, Zheng X**. 2022*a*. Strigolactone alleviates the salinity-alkalinity stress of Malus hupehensis seedlings. Frontiers in Plant Science **13**, 901782.

**Ma Q, Lin X, Zhan M, Chen Z, Wang H, Yao F, Chen J**. 2022*b*. Effect of an exogenous strigolactone GR24 on the antioxidant capacity and quality deterioration in postharvest sweet orange fruit stored at ambient temperature. International Journal of Food Science & Technology **57**, 619–630.

**Manghwar H, Lindsey K, Zhang X, Jin S**. 2019. CRISPR/Cas system: recent advances and future prospects for genome editing. Trends in plant science **24**, 1102–1125.

**Mariotti L, Fambrini M, Pugliesi C, Scartazza A**. 2022. The gibberellin-deficient dwarf2 mutant of sunflower shows a high constitutive level of jasmonic and salicylic acids and an elevated energy dissipation capacity in well-watered and drought conditions. Environmental and Experimental Botany **194**, 104697.

**Martínez-Andújar C, Martínez-Pérez A, Albacete A, *et al.*** 2021. Overproduction of ABA in rootstocks alleviates salinity stress in tomato shoots. Plant, Cell & Environment **44**, 2966–2986.

**Martínez-Medina A, Roldán A, Albacete A, Pascual JA**. 2011. The interaction with arbuscular mycorrhizal fungi or Trichoderma harzianum alters the shoot hormonal profile in melon plants. Phytochemistry **72**, 223–229.

**Marzec M, Daszkowska-Golec A, Collin A, Melzer M, Eggert K, Szarejko I**. 2020. Barley strigolactone signalling mutant hvd14.d reveals the role of strigolactones in abscisic acid-dependent response to drought. Plant, Cell & Environment **43**, 2239–2253.

**Mazorra LM, Holton N, Bishop GJ, Núñez M**. 2011. Heat shock response in tomato brassinosteroid mutants indicates that thermotolerance is independent of brassinosteroid homeostasis. Plant Physiology and Biochemistry **49**, 1420–1428.

**Mekureyaw MF, Pandey C, Hennessy RC, Nicolaisen MH, Liu F, Nybroe O, Roitsch T**. 2022. The cytokinin-producing plant beneficial bacterium Pseudomonas fluorescens G20-18 primes tomato (Solanum lycopersicum) for enhanced drought stress responses. Journal of Plant Physiology **270**, 153629.

**Meyer RS, Purugganan MD**. 2013. Evolution of crop species: genetics of domestication and diversification. Nature reviews genetics **14**, 840–852.

**Mittler R, Zandalinas SI, Fichman Y, Van Breusegem F**. 2022. Reactive oxygen species signalling in plant stress responses. Nature Reviews Molecular Cell Biology 2022 23:10 **23**, 663–679.

**Mizokami Y, Noguchi K, Kojima M, Sakakibara H, Terashima I**. 2015. Mesophyll conductance decreases in the wild type but not in an ABA-deficient mutant (aba1) of *Nicotiana plumbaginifolia* under drought conditions. PLANT CELL AND ENVIRONMENT **38**, 388–398.

**Mohammed AR, Cothren JT, Chen M, Tarpley L**. 2015. 1-Methylcyclopropene (1-MCP)-Induced Alteration in Leaf Photosynthetic Rate, Chlorophyll Fluorescence, Respiration and Membrane Damage in Rice (O ryza sativa L.) Under High Night Temperature. Journal of Agronomy and Crop Science **201**, 105–116.

**Mohi-Ud-Din M, Talukder D, Rohman M, Ahmed JU, Jagadish SVK, Islam T, Hasanuzzaman M**. 2021. Exogenous application of methyl jasmonate and salicylic acid mitigates drought-induced oxidative damages in french bean (Phaseolus vulgaris L.). Plants **10**, 2066.

**Mohsin SM, Hasanuzzaman M, Parvin K, Fujita M**. 2020. Pretreatment of wheat (Triticum aestivum L.) seedlings with 2,4-D improves tolerance to salinity-induced oxidative stress and methylglyoxal toxicity by modulating ion homeostasis, antioxidant defenses, and glyoxalase systems. Plant Physiology and Biochemistry **152**, 221–231.

**Moradi S, Baninasab B, Gholami M, Ghobadi C**. 2017. Paclobutrazol application enhances antioxidant enzyme activities in pomegranate plants affected by cold stress. The Journal of Horticultural Science and Biotechnology **92**, 65–71.

**Mushtaq N, Wang Y, Fan J, Li Y, Ding J**. 2022. Down-Regulation of Cytokinin Receptor Gene SlHK2 Improves Plant Tolerance to Drought, Heat, and Combined Stresses in Tomato. Plants **11**, 154.

**Nadira UA, Ahmed IM, Wu F, Zhang G**. 2016. The regulation of root growth in response to phosphorus deficiency mediated by phytohormones in a Tibetan wild barley accession. Acta Physiologiae Plantarum **38**.

**Naeem M, Basit A, Ahmad I, Mohamed HI, Wasila H**. 2020. Effect of Salicylic Acid and Salinity Stress on the Performance of Tomato Plants. Gesunde Pflanzen **72**.

**Nagar S, Singh VP, Arora A, Dhakar R, Singh N, Singh GP, Meena S, Kumar S, Shiv Ramakrishnan R**. 2021. Understanding the Role of Gibberellic Acid and Paclobutrazol in Terminal Heat Stress Tolerance in Wheat. Frontiers in Plant Science **12**, 692252.

**Najeeb U, Atwell BJ, Bange MP, Tan DKY**. 2015. Aminoethoxyvinylglycine (AVG) ameliorates waterlogging-induced damage in cotton by inhibiting ethylene synthesis and sustaining photosynthetic capacity. Plant Growth Regulation **76**, 83–98.

**Nanda S, Kumar G, Hussain S**. 2021. Utilization of seaweed-based biostimulants in improving plant and soil health: current updates and future prospective. International Journal of Environmental Science and Technology 2021, 1–14.

**Nazara R, Umara S, Khanb NA**. 2015. Exogenous salicylic acid improves photosynthesis and growth through increase in ascorbate-glutathione metabolism and S assimilation in mustard under salt stress. Plant Signaling and Behavior **10**.

**Nitsch L, Kohlen W, Oplaat C, Charnikhova T, Cristescu S, Michieli P, Wolters-Arts M, Bouwmeester H, Mariani C, Vriezen WH**. 2012. ABA-deficiency results in reduced plant and fruit size in tomato. Journal of plant physiology **169**, 878–883.

**Nivetha N, Asha AD, Krishna GK, Chinnusamy V, Paul S**. 2024. Rhizobacteria Bacillus spp. mitigate osmotic stress and improve seed germination in mustard by regulating osmolyte and plant hormone signaling. Physiologia Plantarum **176**.

**Noreen S, Siddiq A, Hussain K, Ahmad S, Hasanuzzama M**. 2017. Foliar application of salicylic acid with salinity stress on physiological and biochemical attributes of sunflower (Helianthus annuus L.) crop. Acta Scientiarum Polonorum. Hortorum Cultus **16**.

**Núñez Vázquez M, Reyes Guerrero Y, Rosabal Ayan L, Martínez L, González Cepero MC**. 2013. Brassinosteroids and its analogs enhance the seedling growth of two rice (Oryza sativa L.) genotypes under saline conditions. *in press*.

**Oguz MC**. 2024. Stimulating endogenous hormone content by plant extracts: increased in vitro regeneration of flax (Linum usitatissimum) cultivars. Journal of Crop Science and Biotechnology, 1–13.

**Okamoto M, Peterson FC, Defries A, Park SY, Endo A, Nambara E, Volkman BF, Cutler SR**. 2013. Activation of dimeric ABA receptors elicits guard cell closure, ABA-regulated gene expression, and drought tolerance. Proceedings of the National Academy of Sciences of the United States of America **110**, 12132–12137.

**Oklestkova J, Rárová L, Kvasnica M, Strnad M**. 2015. Brassinosteroids: synthesis and biological activities. Phytochemistry Reviews 2015 14:6 **14**, 1053–1072.

**Olaetxea M, De Hita D, Garcia CA, *et al.*** 2018. Hypothetical framework integrating the main mechanisms involved in the promoting action of rhizospheric humic substances on plant root- and shoot- growth. Applied Soil Ecology **123**, 521–537.

**Omae N, Tsuda K**. 2022. Plant-Microbiota Interactions in Abiotic Stress Environments. American Phytopathological Society.

**Omena-Garcia RP, Oliveira Martins A, Medeiros DB, Vallarino JG, Mendes Ribeiro D, Fernie AR, Araújo WL, Nunes-Nesi A**. 2019. Growth and metabolic adjustments in response to gibberellin deficiency in drought stressed tomato plants. Environmental and Experimental Botany **159**, 95–107.

**Orozco-Mosqueda M del C, Duan J, DiBernardo M, Zetter E, Campos-García J, Glick BR, Santoyo G**. 2019. The production of ACC deaminase and trehalose by the plant growth promoting bacterium Pseudomonas sp. UW4 synergistically protect tomato plants against salt stress. Frontiers in microbiology **10**, 1392.

**Patel K, Agarwal P, Agarwal PK**. 2018. Kappaphycus alvarezii sap mitigates abiotic-induced stress in Triticum durum by modulating metabolic coordination and improves growth and yield. Journal of Applied Phycology **30**, 2659–2673.

**Paulert R, Ascrizzi R, Malatesta S, *et al.*** 2021. Ulva intestinalis extract acts as biostimulant and modulates metabolites and hormone balance in basil (Ocimum basilicum l.) and parsley (petroselinum crispum l.). Plants **10**.

**Pehlivan N, Yesilyurt AM, Durmus N, Karaoglu SA**. 2017. Trichoderma lixii ID11D seed biopriming mitigates dose dependent salt toxicity in maize. Acta Physiologiae Plantarum **39**.

**Pokotylo I, Hodges M, Kravets V, Ruelland E**. 2022. A ménage à trois: salicylic acid, growth inhibition, and immunity. Trends in Plant Science **27**, 460–471.

**Pospíšilová H, Jiskrova E, Vojta P, Mrizova K, Kokáš F, Čudejková MM, Bergougnoux V, Plíhal O, Klimešová J, Novák O**. 2016. Transgenic barley overexpressing a cytokinin dehydrogenase gene shows greater tolerance to drought stress. New biotechnology **33**, 692–705.

**Pradhan M, Sahoo RK, Swain DM, Dangar TK, Mohanty S**. 2018. Inoculation of Azotobacter vinellandii (SRI Az3) to rice plant increases stress tolerance in rice plant during drought stress. ORYZA-An International Journal on Rice **55**, 406–412.

**Qiao Z, Yao C, Sun S, Zhang F, Yao X, Li X, Zhang J, Jiang X**. 2023. Spraying S-ABA Can Alleviate the Growth Inhibition of Corn (Zea mays L.) Under Water-Deficit Stress. Journal of Soil Science and Plant Nutrition **23**, 1222–1234.

**Qin XQ, Zeevaart JAD**. 2002. Overexpression of a 9-cis-epoxycarotenoid dioxygenase gene in *Nicotiana plumbaginifolia* increases abscisic acid and phaseic acid levels and enhances drought tolerance. PLANT PHYSIOLOGY **128**, 544–551.

**Qiu C-W, Zhang C, Wang N-H, Mao W, Wu F**. 2021. Strigolactone GR24 improves cadmium tolerance by regulating cadmium uptake, nitric oxide signaling and antioxidant metabolism in barley (Hordeum vulgare L.). Environmental Pollution **273**, 116486.

**Rademacher W**. 2015. Plant Growth Regulators: Backgrounds and Uses in Plant Production. Journal of Plant Growth Regulation **34**, 845–872.

**Rakkammal K, Pandian S, Maharajan T, Antony Ceasar S, Sohn SI, Ramesh M**. 2024. Humic acid regulates gene expression and activity of antioxidant enzymes to inhibit the salt-induced oxidative stress in finger millet. Cereal Research Communications **52**, 397–411.

**Raspor M, Mrvaljević M, Savić J, Ćosić T, Kaleri AR, Pokimica N, Cingel A, Ghalawnji N, Motyka V, Ninković S**. 2024. Cytokinin deficiency confers enhanced tolerance to mild, but decreased tolerance to severe salinity stress in in vitro grown potato. Frontiers in Plant Science **14**, 1296520.

**Rathor P, Upadhyay P, Ullah A, Gorim LY, Thilakarathna MS**. 2024. Humic acid improves wheat growth by modulating auxin and cytokinin biosynthesis pathways. AoB PLANTS **16**.

**Rathore RS, Mishra M, Pareek A, Singla-Pareek SL**. 2024. Concurrent improvement of rice grain yield and abiotic stress tolerance by overexpression of cytokinin activating enzyme LONELY GUY (OsLOG). Plant Physiology and Biochemistry **211**, 108635.

**Reyes Guerrero Y, Martínez González L, Núñez Vázquez M**. 2017. Biobras-16 foliar spraying enhances rice (Oryza sativa L.) young plant growth under NaCl treatment. *in press*.

**Rezazadeh A, Harkess RL, Bi G**. 2016. Effects of Paclobutrazol and Flurprimidol on Water Stress Amelioration in Potted Red Firespike. HortTechnology **26**, 26–29.

**Romano I, Ventorino V, Pepe O**. 2020. Effectiveness of Plant Beneficial Microbes: Overview of the Methodological Approaches for the Assessment of Root Colonization and Persistence. Frontiers in Plant Science **11**, 487940.

**Rosheen, Sharma S, Utreja D**. 2023. Salicylic Acid: Synthetic Strategies and Their Biological Activities. ChemistrySelect **8**, e202204614.

**Rouphael Y, du Jardin P, Brown P, De Pascale S, Colla G**. 2020. *Biostimulants for sustainable crop production*. Burleigh Dodds Science Publishing, Cambridge, United Kingdom.

**Ruiz-Lozano JM, Aroca R, Zamarreño ÁM, Molina S, Andreo-Jiménez B, Porcel R, García-Mina JM, Ruyter-Spira C, López-Ráez JA**. 2016. Arbuscular mycorrhizal symbiosis induces strigolactone biosynthesis under drought and improves drought tolerance in lettuce and tomato. Plant Cell and Environment **39**, 441–452.

**Sahni S, Prasad BD, Liu Q, Grbic V, Sharpe A, Singh SP, Krishna P**. 2016. Overexpression of the brassinosteroid biosynthetic gene DWF4 in Brassica napus simultaneously increases seed yield and stress tolerance. Scientific Reports 2016 6:1 **6**, 1–14.

**Sahoo RK, Ansari MW, Pradhan M, Dangar TK, Mohanty S, Tuteja N**. 2014. A novel azotobacter vinellandii (SRIAz3) functions in salinity stress tolerance in rice. Plant Signaling and Behavior **9**.

**Sahoo RK, Rani V, Tuteja N**. 2021. Azotobacter vinelandii helps to combat chromium stress in rice by maintaining antioxidant machinery. 3 Biotech **11**.

**Saleem K, Asghar MA, Raza A, Pan K, Ullah A, Javed HH, Seleiman MF, Imran S, Nadeem SM, Khan KS**. 2024. Alleviating drought stress in strawberry plants: unraveling the role of paclobutrazol as a growth regulator and reducer of oxidative stress induced by reactive oxygen and carbonyl species. Journal of Plant Growth Regulation **43**, 3238–3253.

**Sangha JS, Kelloway S, Critchley AT, Prithiviraj B**. 2014. Seaweeds (Macroalgae) and their extracts as contributors of plant productivity and quality. the current status of our understanding. Advances in Botanical Research. Academic Press Inc., 189–219.

**Saravanakumar D, Samiyappan R**. 2007. ACC deaminase from Pseudomonas fluorescens mediated saline resistance in groundnut (Arachis hypogea) plants. Journal of Applied Microbiology **102**, 1283–1292.

**Sariñana-Aldaco O, Benavides-Mendoza A, Robledo-Olivo A, González-Morales S**. 2022. The Biostimulant Effect of Hydroalcoholic Extracts of Sargassum spp. in Tomato Seedlings under Salt Stress. Plants **11**.

**Saruhan N, Saglam A, Kadioglu A**. 2012. Salicylic acid pretreatment induces drought tolerance and delays leaf rolling by inducing antioxidant systems in maize genotypes. Acta Physiologiae Plantarum **34**, 97–106.

**Serna M, Coll Y, Zapata PJ, Botella MÁ, Pretel MT, Amorós A**. 2015. A brassinosteroid analogue prevented the effect of salt stress on ethylene synthesis and polyamines in lettuce plants. Scientia Horticulturae **185**, 105–112.

**Shahzad R, Khan AL, Bilal S, Waqas M, Kang S-M, Lee I-J**. 2017. Inoculation of abscisic acid-producing endophytic bacteria enhances salinity stress tolerance in Oryza sativa. Environmental and Experimental Botany **136**, 68–77.

**Shakirova FM, Sakhabutdinova AR, Bezrukova M V, Fatkhutdinova RA, Fatkhutdinova DR**. *Changes in the hormonal status of wheat seedlings induced by salicylic acid and salinity*.

**Sharma L, Dalal M, Verma RK, Kumar SV V, Yadav SK, Pushkar S, Kushwaha SR, Bhowmik A, Chinnusamy V**. 2018. Auxin protects spikelet fertility and grain yield under drought and heat stresses in rice. Environmental and Experimental Botany **150**, 9–24.

**Sheikh Mohammadi MH, Etemadi N, Arab MM, Aalifar M, Arab M, Pessarakli M**. 2017. Molecular and physiological responses of Iranian Perennial ryegrass as affected by Trinexapac ethyl, Paclobutrazol and Abscisic acid under drought stress. Plant Physiology and Biochemistry **111**, 129–143.

**Sheteiwy MS, Dong Q, An J, Song W, Guan Y, He F, Huang Y, Hu J**. 2017. Regulation of ZnO nanoparticles-induced physiological and molecular changes by seed priming with humic acid in Oryza sativa seedlings. Plant Growth Regulation **83**, 27–41.

**Shukry WM, Abu-Ria ME, Abo-Hamed SA, Anis GB, Ibraheem F**. 2023. The Efficiency of Humic Acid for Improving Salinity Tolerance in Salt Sensitive Rice (Oryza sativa): Growth Responses and Physiological Mechanisms. Gesunde Pflanzen **75**, 2639–2653.

**Siddiqui MH, Khan MN, Mohammad F, Khan MMA**. 2008. Role of nitrogen and gibberellin (GA3) in the regulation of enzyme activities and in osmoprotectant accumulation in Brassica juncea L. under salt stress. Journal of Agronomy and Crop Science **194**, 214–224.

**Singh RP, Pandey DM, Jha PN, Ma Y**. 2022. ACC deaminase producing rhizobacterium Enterobacter cloacae ZNP-4 enhance abiotic stress tolerance in wheat plant. PLoS ONE **17**.

**Singh S, Tripathi A, Chanotiya CS, Barnawal D, Singh P, Patel VK, Vajpayee P, Kalra A**. 2020. Cold stress alleviation using individual and combined inoculation of ACC deaminase producing microbes in Ocimum sanctum. Environmental Sustainability **3**, 289–301.

**Sitohy MZ, Desoky ESM, Osman A, Rady MM**. 2020. Pumpkin seed protein hydrolysate treatment alleviates salt stress effects on Phaseolus vulgaris by elevating antioxidant capacity and recovering ion homeostasis. Scientia Horticulturae **271**.

**Song Y, Lv D, Jiang M, E Z, Han Y, Sun Y, Zhu S, Chen J, Zhao T**. 2023. Exogenous strigolactones enhance salinity tolerance in cotton (Gossypium hirsutum L.) seedlings. Plant Stress **10**, 100235.

**Sorrentino M, De Diego N, Ugena L, Spíchal L, Lucini L, Miras-Moreno B, Zhang L, Rouphael Y, Colla G, Panzarová K**. 2021. Seed Priming With Protein Hydrolysates Improves Arabidopsis Growth and Stress Tolerance to Abiotic Stresses. Frontiers in Plant Science **12**.

**Soskine M, Tawfik DS**. 2010. Mutational effects and the evolution of new protein functions. Nature Reviews Genetics **11**, 572–582.

**Sulpice R, McKeown PC**. 2015. Moving toward a comprehensive map of central plant metabolism. Annual Review of Plant Biology **66**, 187–210.

**Sun X, Hu C, Tan Q, Liu J, Liu H**. 2009. Effects of molybdenum on expression of cold-responsive genes in abscisic acid (ABA)-dependent and ABA-independent pathways in winter wheat under low-temperature stress. Annals of Botany **104**, 345–356.

**Sun L, Zhang Q, Wu J, Zhang L, Jiao X, Zhang S, Zhang Z, Sun D, Lu T, Sun Y**. 2014. Two Rice Authentic Histidine Phosphotransfer Proteins, OsAHP1 and OsAHP2, Mediate Cytokinin Signaling and Stress Responses in Rice. Plant Physiology **165**, 335–345.

**Takaoka Y, Iwahashi M, Chini A, *et al.*** 2018. A rationally designed JAZ subtype-selective agonist of jasmonate perception. Nature Communications **9**, 3654.

**Tanaka N, Matsuoka M, Kitano H, Asano T, Kaku H, Komatsu S**. 2006. gid1, a gibberellin-insensitive dwarf mutant, shows altered regulation of probenazole-inducible protein (PBZ1) in response to cold stress and pathogen attack. Plant, Cell & Environment **29**, 619–631.

**Thakur R, Yadav S**. 2024. Biofilm forming, exopolysaccharide producing and halotolerant, bacterial consortium mitigates salinity stress in Triticum aestivum. International Journal of Biological Macromolecules **262**.

**Tian Y, Zhang H, Pan X, Chen X, Zhang Z, Lu X, Huang R**. 2011. Overexpression of ethylene response factor TERF2 confers cold tolerance in rice seedlings. Transgenic research **20**, 857–866.

**Timm CM, Carter KR, Carrell AA, *et al.*** 2018. Abiotic Stresses Shift Belowground Populus -Associated Bacteria Toward a Core Stress Microbiome . mSystems **3**.

**Torun H, Novák O, Mikulík J, Strnad M, Ayaz FA**. 2022. The Effects of Exogenous Salicylic Acid on Endogenous Phytohormone Status in Hordeum vulgare L. under Salt Stress. Plants **11**.

**Tounekti T, Hernández I, Müller M, Khemira H, Munné-Bosch S**. 2011. Kinetin applications alleviate salt stress and improve the antioxidant composition of leaf extracts in Salvia officinalis. Plant Physiology and Biochemistry **49**, 1165–1176.

**Trevisan S, Manoli A, Quaggiotti S**. 2019. A novel biostimulant, belonging to protein hydrolysates, mitigates abiotic stress effects on maize seedlings grown in hydroponics. Agronomy **9**.

**Tripathi D, Raikhy G, Kumar D**. 2019. Chemical elicitors of systemic acquired resistance—Salicylic acid and its functional analogs. Current Plant Biology **17**, 48–59.

**Turkan I**. 2018. ROS and RNS: key signalling molecules in plants. Journal of Experimental Botany **69**, 3313–3315.

**Tworkoski T, Wisniewski M, Artlip T**. 2011. Application of BABA and s-ABA for drought resistance in apple. United States Department of Agriculture-Agricultural Research Service / University of Nebraska-Lincoln: Faculty Publications *in press*.

**Ullah A, Zeng F, Tariq A, Asghar MA, Saleem K, Raza A, Naseer MA, Zhang Z, Noor J**. 2022. Exogenous naphthaleneacetic acid alleviated alkalinity-induced morpho-physio-biochemical damages in Cyperus esculentus L. var. sativus Boeck. Frontiers in Plant Science **13**, 1018787.

**Vaidya AS, Helander JDM, Peterson FC, *et al.*** 2019. Dynamic control of plant water use using designed ABA receptor agonists. Science **366**, eaaw8848.

**Visentin I, Vitali M, Ferrero M, Zhang Y, Ruyter-Spira C, Novák O, Strnad M, Lovisolo C, Schubert A, Cardinale F**. 2016. Low levels of strigolactones in roots as a component of the systemic signal of drought stress in tomato. New Phytologist **212**, 954–963.

**WALIA US.** 2023. MODERN TECHNIQUES FOR CROP MANAGEMENT. *in press*.

**Walker-Simmons M, Kudrna DA, Warner RL**. 1989. Reduced accumulation of ABA during water stress in a molybdenum cofactor mutant of barley. Plant Physiology **90**, 728–733.

**Wang Y, Hu J, Qin G, Cui H, Wang Q**. 2012. Salicylic acid analogues with biological activity may induce chilling tolerance of maize (Zea mays) seeds. Botany **90**, 845–855.

**Wang M, Qiao J, Yu C, Chen H, Sun C, Huang L, Li C, Geisler M, Qian Q, Jiang DA**. 2019*a*. The auxin influx carrier, OsAUX3, regulates rice root development and responses to aluminium stress. Plant, cell & environment **42**, 1125–1138.

**Wang F, Wan C, Wu W, Pan Y, Cheng X, Li C, Pi J, Chen X**. 2023*a*. Exogenous methyl jasmonate (MeJA) enhances the tolerance to cadmium (Cd) stress of okra (Abelmoschus esculentus L.) plants. Plant Cell, Tissue and Organ Culture (PCTOC) **155**, 907–922.

**Wang F, Wan C, Wu W, Zhang Y, Pan Y, Chen X, Li C, Pi J, Wang Z, Ye Y**. 2023*b*. Methyl jasmonate (MeJA) enhances salt tolerance of okra (Abelmoschus esculentus L.) plants by regulating ABA signaling, osmotic adjustment substances, photosynthesis and ROS metabolism. Scientia Horticulturae **319**, 112145.

**Wang N, Wang X, Shi J, Liu X, Xu Q, Zhou H, Song M, Yan G**. 2019*b*. Mepiquat chloride-priming induced salt tolerance during seed germination of cotton (Gossypium hirsutum L.) through regulating water transport and K+/Na+ homeostasis. Environmental and Experimental Botany **159**, 168–178.

**Wang J, Wang D, Zhu M, Li F**. 2021. Exogenous 6-benzyladenine improves waterlogging tolerance in maize seedlings by mitigating oxidative stress and upregulating the ascorbate-glutathione cycle. Frontiers in Plant Science **12**, 680376.

**Wang X, Xu C, Cang J, Zeng Y, Yu J, Liu L, Zhang D, Wang J**. 2015. *Effects of Exogenous GA 3 on Wheat Cold Tolerance*.

**Wang D, Yang Z, Wu M, Wang W, Wang Y, Nie S**. 2022*a*. Enhanced brassinosteroid signaling via the overexpression of SlBRI1 positively regulates the chilling stress tolerance of tomato. Plant Science **320**, 111281.

**Wang C, Yang A, Yin H, Zhang J**. 2008. Influence of water stress on endogenous hormone contents and cell damage of maize seedlings. Journal of Integrative Plant Biology **50**, 427–434.

**Wang W, Zhang C, Zheng W, Lv H, Li J, Liang B, Zhou W**. 2022*b*. Seed priming with protein hydrolysate promotes seed germination via reserve mobilization, osmolyte accumulation and antioxidant systems under PEG-induced drought stress. Plant Cell Reports **41**, 2173–2186.

**Weijers D, Nemhauser J, Yang Z**. 2018. Auxin: small molecule, big impact. Journal of Experimental Botany **69**, 133–136.

**Wi SJ, Jang SJ, Park KY**. 2010. Inhibition of biphasic ethylene production enhances tolerance to abiotic stress by reducing the accumulation of reactive oxygen species in Nicotiana tabacum. Molecules and Cells **30**, 37–49.

**Win NM, Yoo J, Lwin HP, Lee EJ, Kang IK, Lee J**. 2021. Effects of 1-methylcyclopropene and aminoethoxyvinylglycine treatments on fruit quality and antioxidant metabolites in cold-stored 'Sangjudungsi' persimmons. Horticulture Environment and Biotechnology **62**, 891–905.

**WMO**. 2024. Climate change indicators reached record levels in 2023: WMO. https://wmo.int/news/media-centre/climate-change-indicators-reached-record-levels-2023-wmo. Accessed August 2024.

**Wu H, Bai B, Lu X, Li H**. 2023. A gibberellin-deficient maize mutant exhibits altered plant height, stem strength and drought tolerance. Plant Cell Reports **42**, 1687–1699.

**Wu H, Wu X, Li Z, Duan L, Zhang M**. 2012. Physiological evaluation of drought stress tolerance and recovery in cauliflower (Brassica oleracea L.) seedlings treated with methyl jasmonate and coronatine. Journal of Plant Growth Regulation **31**, 113–123.

**Wu K, Xu H, Gao X, Fu X**. 2021. New insights into gibberellin signaling in regulating plant growth–metabolic coordination. Current Opinion in Plant Biology **63**, 102074.

**Wu W, Zhang Q, Ervin EH, Yang Z, Zhang X**. 2017. Physiological mechanism of enhancing salt stress tolerance of perennial ryegrass by 24-epibrassinolide. Frontiers in plant science **8**, 1017.

**Xing X, Jiang H, Zhou Q, Xing H, Jiang H, Wang S**. 2016*a*. Improved drought tolerance by early IAA-and ABA-dependent H 2 O 2 accumulation induced by α-naphthaleneacetic acid in soybean plants. Plant Growth Regulation **80**, 303–314.

**Xing X, Jiang H, Zhou Q, Xing H, Jiang H, Wang S**. 2016*b*. Improved drought tolerance by early IAA- and ABA-dependent H2O2 accumulation induced by α-naphthaleneacetic acid in soybean plants. Plant Growth Regulation **80**, 303–314.

**Xu Y, Burgess P, Zhang X, Huang B**. 2016. Enhancing cytokinin synthesis by overexpressing ipt alleviated drought inhibition of root growth through activating ROS-scavenging systems in Agrostis stolonifera. Journal of Experimental Botany **67**, 1979–1992.

**Xu Y, Tian J, Gianfagna T, Huang B**. 2009. Effects of SAG12-ipt expression on cytokinin production, growth and senescence of creeping bentgrass (Agrostis stolonifera L.) under heat stress. Plant Growth Regulation **57**, 281–291.

**Yadav G, Sharma N, Goel A, Varma A, Mishra A, Kothari SL, Choudhary DK**. 2024. Trichoderma Mediated Metal Chelator and Its Role in Solanum melongena Growth Under Heavy Metals. Journal of Plant Growth Regulation **43**, 178–200.

**Yaghoubi Khanghahi M, AbdElgawad H, Verbruggen E, Korany SM, Alsherif EA, Beemster GTS, Crecchio C**. 2022. Biofertilisation with a consortium of growth-promoting bacterial strains improves the nutritional status of wheat grain under control, drought, and salinity stress conditions. Physiologia Plantarum **174**.

**Yan J, Li S, Gu M, *et al.*** 2016. Endogenous Bioactive Jasmonate Is Composed of a Set of (+)-7-iso-JA-Amino Acid Conjugates. Plant Physiology **172**, 2154–2164.

**Yang D, Li Y, Shi Y, Cui Z, Luo Y, Zheng M, Chen J, Li Y, Yin Y, Wang Z**. 2016. Exogenous cytokinins increase grain yield of winter wheat cultivars by improving stay-green characteristics under heat stress. PLoS One **11**, e0155437.

**Yang D, Luo Y, Ni Y, Yin Y, Yang W, Peng D, Cui Z, Wang Z**. 2014. Effects of exogenous ABA application on post-anthesis dry matter redistribution and grain starch accumulation of winter wheat with different staygreen characteristics. Crop Journal **2**, 144–153.

**Yang C, Wang H, Ouyang Q, Chen G, Fu X, Hou D, Xu H**. 2023. Deficiency of Auxin Efflux Carrier OsPIN1b Impairs Chilling and Drought Tolerance in Rice. Plants **12**, 4058.

**Yao C, Zhang F, Sun X, Shang D, He F, Li X, Zhang J, Jiang X**. 2019. Effects of S-Abscisic Acid (S-ABA) on Seed Germination, Seedling Growth, and Asr1 Gene Expression Under Drought Stress in Maize. Journal of Plant Growth Regulation **38**, 1300–1313.

**Ye L, Zhao X, Bao E, Cao K, Zou Z**. 2019. Effects of arbuscular mycorrhizal fungi on watermelon growth, elemental uptake, antioxidant, and photosystem ii activities and stress-response gene expressions under salinity-alkalinity stresses. Frontiers in Plant Science **10**.

**Yu C, Sun C, Shen C, Wang S, Liu F, Liu Y, Chen Y, Li C, Qian Q, Aryal B**. 2015. The auxin transporter, Os AUX 1, is involved in primary root and root hair elongation and in Cd stress responses in rice (Oryza sativa L.). The Plant Journal **83**, 818–830.

**Zahedi H, Abbasi S**. 2015. Effect of plant growth promoting rhizobacteria (PGPR) and water stress on phytohormones and polyamines of soybean. Indian Journal of Agricultural Research **49**, 427–431.

**Zhang Z, Li F, Li D, Zhang H, Huang R**. 2010. Expression of ethylene response factor JERF1 in rice improves tolerance to drought. Planta **232**, 765–774.

**Zhang X-H, Ma C, Zhang L, Su M, Wang J, Zheng S, Zhang T-G**. 2022. GR24-mediated enhancement of salt tolerance and roles of H2O2 and Ca2+ in regulating this enhancement in cucumber. Journal of Plant Physiology **270**, 153640.

**Zhang W, Wang J, Huang Z, Mi L, Xu K, Wu J, Fan Y, Ma S, Jiang D**. 2019. Effects of low temperature at booting stage on sucrose metabolism and endogenous hormone contents in winter wheat spikelet. Frontiers in Plant Science **10**.

**Zhang H, Zhu J, Gong Z, Zhu JK**. 2021. Abiotic stress responses in plants. Nature Reviews Genetics 2021 23:2 **23**, 104–119.

**Zhao Y**. 2018. Essential roles of local auxin biosynthesis in plant development and in adaptation to environmental changes. Annual Review of Plant Biology **69**, 417–435.

**Zhao S, Ma Q, Xu X, Li G, Hao L**. 2016. Tomato Jasmonic Acid-Deficient Mutant spr2 Seedling Response to Cadmium Stress. Journal of Plant Growth Regulation **35**, 603–610.

**Zhou L, Monreal CM, Xu S, McLaughlin NB, Zhang H, Hao G, Liu J**. 2019. Effect of bentonite-humic acid application on the improvement of soil structure and maize yield in a sandy soil of a semi-arid region. Geoderma **338**, 269–280.

**Zou P, Lu X, Zhao H, Yuan Y, Meng L, Zhang C, Li Y**. 2019. Polysaccharides derived from the brown algae Lessonia nigrescens enhance salt stress tolerance to wheat seedlings by enhancing the antioxidant system and modulating intracellular ion concentration. Frontiers in Plant Science **10**.

**Zuluaga MYA, Monterisi S, Rouphael Y, Colla G, Lucini L, Cesco S, Pii Y**. 2023. Different vegetal protein hydrolysates distinctively alleviate salinity stress in vegetable crops: A case study on tomato and lettuce. Frontiers in Plant Science **14**.

**Zwack PJ, Rashotte AM**. 2015. Interactions between cytokinin signalling and abiotic stress responses. Journal of Experimental Botany **66**, 4863–4871.

**Zwanenburg B, Blanco-Ania D**. 2018. Strigolactones: new plant hormones in the spotlight. Journal of Experimental Botany **69**, 2205–2218.